\PassOptionsToPackage{square,numbers}{natbib} 
\documentclass[10pt]{article}
\usepackage[utf8]{inputenc}
\usepackage[T1]{fontenc}
\usepackage[english]{babel}
\usepackage{graphicx}
\usepackage{amsmath,amssymb,amsbsy,amsthm,color,natbib}
\usepackage{multicol}
\usepackage{color} 
\usepackage{natbib}
\usepackage[linktoc=all,colorlinks=true,citecolor=blue,linkcolor=blue]{hyperref}
\usepackage[paperwidth=18cm,paperheight=29.7cm,hmargin=2.5cm,vmargin=2.5cm]{geometry}
\usepackage{authblk}
\usepackage{comment}
\usepackage{cleveref}
\usepackage{subcaption}
\graphicspath{{figures/}}
\crefformat{equation}{#2{\color{red}#1}#3}
\Crefformat{equation}{#2{\itshape\color{red}Equation~#1}#3}
\crefformat{figure}{#2{\color{cyan}Fig.~#1}#3}
\crefformat{multifigures}{{#2{\color{cyan}Figs.~#1}#3}}
\Crefformat{figure}{#2{\color{cyan}Figure~#1}#3}
\Crefformat{multifigures}{{#2{\color{cyan}Figures~#1}#3}}
\crefformat{section}{#2{\color{cyan}Sect.~#1}#3}
\Crefformat{section}{#2{\color{cyan}Section~#1}#3}
\crefformat{chapter}{#2{\color{cyan}Chapt.~#1}#3}
\Crefformat{chapter}{#2{\color{cyan}Chapter~#1}#3}
\crefformat{table}{#2{\color{cyan}Table~#1}#3}
\Crefformat{table}{#2{\color{cyan}Table~#1}#3}

\title{Manifold dynamics and orbital properties in a two-dimensional galactic model}

\author{Dylan Theron\thanks{Email: dylantheron1907@gmail.com}}
\author{Charalampos Skokos\thanks{Email: haris.skokos@uct.ac.za}}

\affil{Nonlinear Dynamics and Chaos Group \\ Department of Mathematics and Applied Mathematics \\ University of Cape Town, Rondebosch 7701, South Africa}

\begin{document}

\maketitle

\begin{abstract}
We numerically investigate the orbital dynamics of a two-dimensional galactic model, emphasizing the influence of stable and unstable manifolds on the evolution of orbits. In our analysis we use evaluations of the system's Lagrangian descriptors to reveal the structure and location of  manifolds. In addition, we perform extensive computations of the forward and backward in time evolution of ensembles of orbits, which allow us to study the future and past dynamics of the model by constructing its so-called origin-fate maps.  Focusing on the properties of the escape of orbits from the central regions of the model's configuration space, we analyze various aspects of the orbits' evolutions, like the time an orbit needs to escape from the central region, the total time it spends at the exterior areas of the galaxy, as well as the number of its escapes and reentries from and to the central region. Furthermore, we follow the past and future evolution of orbits keeping track of the regions from which they leave  the central parts of the galaxy, and relate different orbital behaviors with the enhancement of specific morphological features in the system's configuration space. Our results indicate that the stable (unstable) manifolds  mainly influence the orbital characteristics of the model in the future (past).
\end{abstract}

\section{Introduction}
\label{sec:intro}

Disk galaxies can exhibit a number of interesting morphological structures like bars \cite{MJ07,BJM08}, spiral arms \cite{EE82,EFPQDDHKRSTT00}, and different types of rings (nuclear rings near the nucleus of the galaxy, inner rings surrounding the bar, or outer rings appearing at larger distances from the center of the galaxy) \cite{B86,B95}. Orbital dynamics provides information about the creation and properties of such structures, by identifying the types of stellar orbits in galaxy models which support the formation of these morphologies \cite{A84,SW93}. Stable periodic orbits are surrounded by  regular orbits and thus constitute the backbone of galaxy structure \cite{ABMP83}.Consequently, the appearance of a specific morphological feature can often be associated with the properties of one of the main families of periodic orbits of the galactic model, like the so-called `x1' family, which supports the creation of bars in a two-dimensional (2D) system \cite{CG89}. It has been shown that in three-dimensional (3D) galactic bar potentials, the backbone of the orbital structure is not just the x1 family, but a tree of 2D and 3D families bifurcating from it \cite{SPA02a,SPA02b,PSA02,PSA03}. Nowadays, it is known that not only regular, but also chaotic sticky orbits can support the formation of various galactic morphological features like bars \cite{KC96,WP99,CH13,TP15}. Furthermore, in \cite{RMAG06,RAMG07,ARM09,ARBM09} a mechanism based on orbital motion driven by the invariant manifolds associated to the periodic orbits around the unstable equilibrium points of a rotating bar potential, was proposed to explain the formation of spirals and rings in barred galaxies. Thus, investigating the orbital dynamics of galactic potentials is an active and vibrant research field, which has already provided many important results in the field of galactic dynamics. 

A  numerical technique which efficiently allows the identification of phase space structures, like the presence of invariant manifolds, in dynamical systems, is the method of Lagrangian descriptors (LDs)~\cite{MM09,MWCM13,AAGGKKNW20}. The values of LDs are obtained through the time integration, over a finite time interval both forward and backward in time,  of a positive quantity along the path of individual orbits. The evaluation of LDs on a  set of initial conditions produces a scalar field which provides a comprehensive picture of the system's dynamics, while singular features in the gradient of this field denote the location of invariant manifolds. The method of Lagrangian descriptors was first introduced in \cite{MM09} and was initially applied to the study of ocean currents~\cite{MM09,MWCM13}. Over the years LDs have been successfully used in studies of dynamical systems from diverse scientific fields such as chemical reaction and molecular structure dynamics \cite{CH15,JH16,CJH17,RBB19,KGW20,AGKW21,RABB23,KHSW23}, as well as geophysical \cite{MM10,GRMCW16,BCGM23}, cardiovascular \cite{DNDK21}, and biomedical flows \cite{AGAM23}. Recently quantities based on LD computations have  been used to identify chaos in different dynamical systems \cite{DPAGM22,HZNKWS22,ZNHKWS23,DC23,CDL25}. 

In this paper we use the method of LDs to reveal the structure of invariant manifolds in a simple 2D  galactic model, and we investigate the influence of these manifolds on the system's orbital behavior. The dynamical model we study has been previously used in investigations of generic phase space structures appearing in the vicinity of different types of stable and unstable periodic orbits \cite{KP11,KPC11,KPC13}. In our study we also use the so-called origin fate map (OFM) \cite{HKWS23}, which allows the clear identification of phase  space regions where orbits with different dynamical behaviors in their past and  future evolution exist.

The paper is organized as follows. In Sect.~\ref{sec:Ham} we describe in detail the Hamiltonian model we use in our investigation, while in Sect.~\ref{sec:LDs} we provide information about the method of LDs, as well as on how we can use this approach  to identify the location of the system's stable and unstable manifolds. Section \ref{sec:Total_effect} contains the numerical results of our investigation. There we study the effect of the manifolds on various aspects of the evolution of ensembles of orbits, discussing also their connections with specific morphological features in the configuration space of the galactic model. Finally, in Sect.~\ref{sec:summary}, we summarize the findings of our work.

\section{The 2D Hamiltonian model}
\label{sec:Ham}

The 2D Hamiltonian model we consider describes the motion of a star in the $(x,y)$ equatorial plane of a 3D galactic system comprised of two Miyamoto disks \cite{MN75}, so that the third spatial coordinate $z$ is always set $z=0$. The 3D version of the model was extensively used in studies of phase space structures near different types of periodic orbits \cite{KP11,KPC11,KPC13}. In particular, the system we use in our study rotates around the  $z$ axis with angular velocity $\Omega_{\mathrm{b}}$, which is also referred to as the system's  \textit{pattern speed}. The Hamiltonian function of the model is 
\begin{equation}
	H(x,y,\dot{x},\dot{y})=\frac{1}{2}(\dot{x}^2+\dot{y}^2)+\Phi(x,y)-\frac{1}{2}\Omega_{\mathrm{b}}^2(x^2+y^2),
	\label{eq:2DHam}
\end{equation}
where $x$, $y$ are the Cartesian coordinates of the star, and  $\dot{x}$, $\dot{y}$ are  their  time derivatives (velocities). In our study we refer to  the numerical value of the Hamiltonian as the system's \textit{energy} $E$. The considered 2D potential $\Phi(x,y)$ is derived as the restriction of the general 3D potential $\Phi_3(x,y,z)$ of two Minamoto disks of  masses $M_1$ and $M_2$
\begin{equation}
	\Phi_3(x,y,z)=-\frac{GM_1}{\sqrt{x^2+\frac{y^2}{q_{\mathrm{a}}^2}+\left(a_1+\sqrt{\frac{z^2}{q_{\mathrm{b}}^2}+b_1^2}\right)^2}}-\frac{GM_2}{\sqrt{x^2+\frac{y^2}{q_{\mathrm{a}}^2}+\left(a_2+\sqrt{\frac{z^2}{q_{\mathrm{b}}^2}+b_2^2}\right)^2}},
	\label{eq:3Dpot}
\end{equation}
on the $(x,y)$ equatorial plane by setting $z=0$ in \eqref{eq:3Dpot}, and
has the form
\begin{equation}
	\Phi(x,y)=-\frac{GM_1}{\sqrt{x^2+\frac{y^2}{q_{\mathrm{a}}^2}+(a_1+b_1)^2}}-\frac{GM_2}{\sqrt{x^2+\frac{y^2}{q_{\mathrm{a}}^2}+(a_2+b_2)^2}}.
	\label{eq:2Dpot}
\end{equation}
We note that the axisymmetric form ($q_{\mathrm{a}} = q_{\mathrm{b}} = 1$) of \eqref{eq:3Dpot} can be considered as an approximation of the potential of the Milky Way \cite{MN75}. 

In our study we follow \cite{KP11} in setting the parameter values and units of system \eqref{eq:2DHam}. In particular, the mass unit $m_{\mathrm{u}}$ corresponds to $10^{10} M_{\odot}$, with $M_{\odot}$ being the solar mass, so that  $m_{\mathrm{u}} =1.989\cdot10^{30}\mbox{kg}$, the distance unit $r_{\mathrm{u}}$ is  1 kpc, and the time unit $t_{\mathrm{u}}$ is approximately $1.4718889907\cdot10^{14} \mbox{s}$. These choices  set the gravitational constant to $G \approx 1$. Following \cite{KP11} we set the parameter values of the system to $M_1=2.05 \times 10^{10} M_{\odot}$, $a_1=0$ kpc, $b_1=0.495$ kpc, $M_2=25.47 \times 10^{10} M_{\odot}$, $a_2=7.258$ kpc, $b_2=0.520$ kpc, $q_{\mathrm{a}}=1.2$, and consider various values  for the pattern speed $\Omega_{\mathrm{b}}$. We note that in \cite{KP11} $q_{\mathrm{b}}=0.9$ but since in our case we consider the restricted version of potential $\Phi_3(x,y,z)$ \eqref{eq:3Dpot} for $z=0$, $q_{\mathrm{b}}$ does not appear in $\Phi(x,y)$ \eqref{eq:2Dpot}.

\begin{figure}[tb!]
	\centering
		\includegraphics[width=\textwidth]{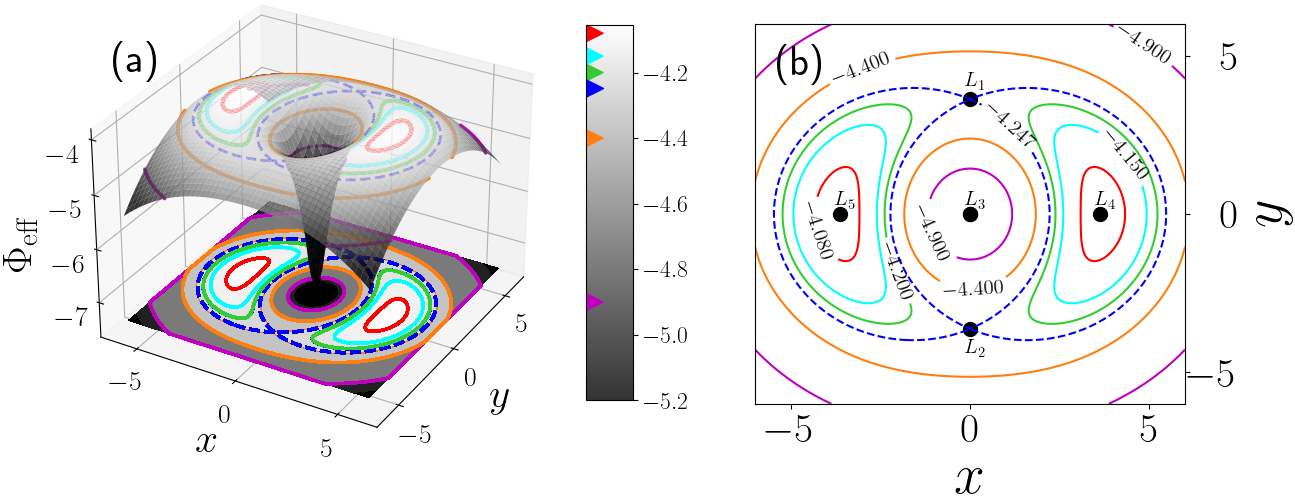}    
	\caption{(a) The 3D  energy surface of the effective potential $\Phi_{\mathrm{e}}(x,y)$ \eqref{eq:effpot} for $\Omega_{\mathrm{b}}=60\,\mbox{km} \,\mbox{s}^{-1}\, \mbox{kpc}^{-1}$, along with its 2D projection on the $(x,y)$ plane (floor of the plot), where points are colored according to the gray color scale at the right side of the panel. Contours corresponding to $\Phi_{\mathrm{e}}(x,y)$ values -4.9 (magenta curves), -4.4 (orange curves), -4.247 (blue dashed curves), -4.2 (green curves), -4.15 (cyan curves), and $-4.08$ (red curves) are shown on both the 3D surface and its 2D projection, as well as in the contour plot of panel (b). These values are also indicated by appropriately colored triangles on the color bar of  (a).  In (b) the blue dashed curves pass through  equilibrium points $L_1$ and $L_2$ (denoted black dots),  the black dot at $(x,y)=(0,0)$  denotes $L_3$, while the remaining two dots represent  equilibrium points $L_4$ and $L_5$. 
	}
	\label{fig:effpot}
\end{figure}

Hamiltonian \eqref{eq:2DHam} can be written as the sum of the kinetic energy  $(\dot{x}^2+\dot{y}^2)/2$ and the effective potential 
\begin{equation}
	\Phi_{\mathrm{e}}(x,y)=-\frac{GM_1}{\sqrt{x^2+\frac{y^2}{q_{\mathrm{a}}^2}+(a_1+b_1)^2}}-\frac{GM_2}{\sqrt{x^2+\frac{y^2}{q_{\mathrm{a}}^2}+(a_2+b_2)^2}}-\frac{1}{2}\Omega_{\mathrm{b}}^2(x^2+y^2).
	\label{eq:effpot}
\end{equation}
The geometry of the energy surface of the effective potential $\Phi_{\mathrm{e}}(x,y)$,  depicted in Fig.~\ref{fig:effpot} for $\Omega_{\mathrm{b}}=60\,\mbox{km} \,\mbox{s}^{-1}\, \mbox{kpc}^{-1}$, allows us to better understand the system's orbital dynamics. In Fig.~\ref{fig:effpot}(a) the 3D $\Phi_{\mathrm{e}}(x,y)$ surface is colored according to the gray color scale at  the right side of the panel. On this surface we have also plotted six contour curves corresponding to  $\Phi_{\mathrm{e}}(x,y)=$ -4.9 (magenta curves), -4.4 (orange curves), -4.247 (blue dashed curves), -4.2 (green curves), -4.15 (cyan curves), and $-4.08$ (red curves). The 3D surface along with the contour curves have  been projected onto the floor of Fig.~\ref{fig:effpot}(a), while the contours are also plotted in Fig.~\ref{fig:effpot}(b). 

The effective potential \eqref{eq:effpot} is characterized by five stationary points, the so-called \textit{Lagrange points} $L_1$ to $L_5$, at which $	\frac{\partial\Phi_{\mathrm{e}}}{\partial x}=\frac{\partial\Phi_{\mathrm{e}}}{\partial y}=0$. These points are denoted by black dots in Fig.~\ref{fig:effpot}(b). The central stationary point $L_3$  is a minimum of the effective potential. Positioning a star at each of the four points $L_1$, $L_2$, $L_4$, and $L_5$, results to a circular orbit in the configuration space $(x,y)$, which corresponds to a stationary point in the $\Omega_{\mathrm{b}}$ rotating frame. The stationary points $L_1$ and $L_2$, located at the intersections of the dashed blue curves of Fig.~\ref{fig:effpot}, are saddle points, while the stationary points $L_4$ and $L_5$ along the $y=0$ axis are maxima of the effective potential.

The energy value of the $L_1$ and $L_2$ points separate regions of higher energies, which are denoted by lighter gray color colors in Fig.~\ref{fig:effpot}(a) containing crescent shaped contours, from lower energy values in  darker colored areas close to the origin $(0,0)$ or far away from it, where circular shaped contours exist.  A star with energy slightly larger than  that of the $L_1$ and $L_2$ points, but less than the energy of the $L_4$ and $L_5$ points, has the  capability to move between the region near the origin, which we will refer to as the \textit{central region}, and the lower energy areas further away from the origin, which we call the \textit{exterior region}. However, the star does not have sufficient energy to visit the so-called  \textit{energetically forbidden regions}, having crescent-like shapes in the vicinity of points $L_4$ and $L_5$. In this case, the star can travel between the central region and the exterior region through the channels between the energetically forbidden regions, a transition which we call \textit{escape}. It was claimed in \cite{D65} that orbits which escape from the interior to the exterior region through the openings between the energetically forbidden areas, contribute to the ring and spiral structure of galaxies. Such orbital behaviors, influenced by the invariant manifolds associated with the periodic orbits (the so-called \textit{Lyapunov orbits})  around the  $L_1$ and $L_2$ points, were studied in detail in \cite{RMAG06,RAMG07,ARM09,ARBM09}. Obviously, stars with energies smaller than that of the $L_1$ and $L_2$ points, can either exist at the central  or the exterior region, but do not have enough energy to move between these regions.  On the other hand, stars with energies  greater than that of the $L_4$ and $L_5$ points do not see any energetically forbidden areas and  may travel to any spatial region. 

\begin{figure}[tb!]
	\centering
		\includegraphics[width=\textwidth]{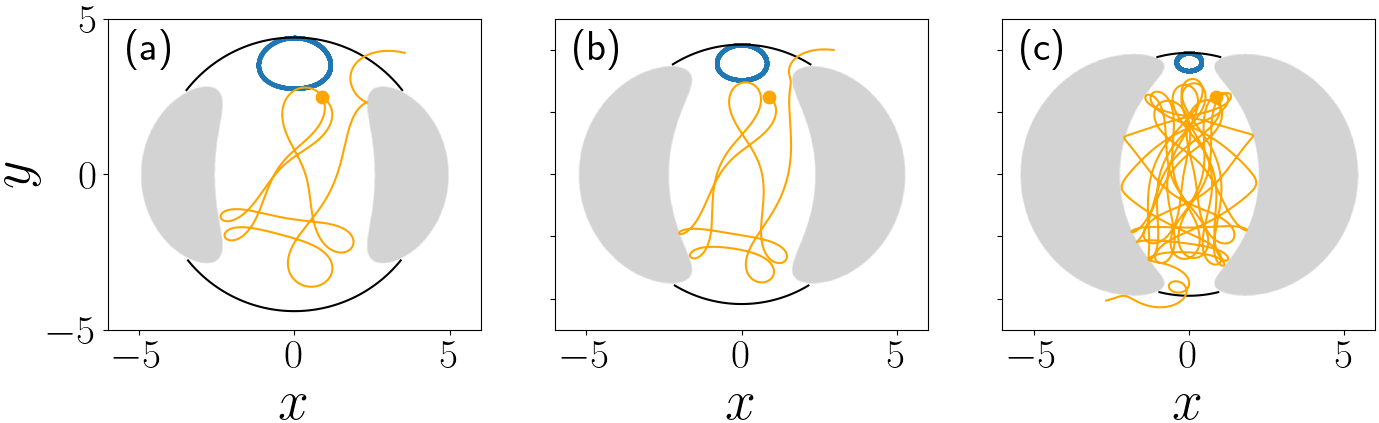}   
	\caption{The configuration space projection  of orbits (orange curves) of the Hamiltonian system \eqref{eq:2DHam} with $\Omega_{\mathrm{b}}=60\,\mbox{km} \,\mbox{s}^{-1}\, \mbox{kpc}^{-1}$, starting at point $(x,y)=(0.88, 2.48)$ (indicated by an orange dot) with $\dot{y}=0$ and $\dot{x}\geq 0$ for energy (a) $E=-4.15$, (b) $E=-4.2$ and (c) $E=-4.235$.  In each panel, the energetically forbidden regions are colored in gray, the Lyapunov orbit at the $y>0$ opening is plotted in blue (a Lyapunov orbit symmetric to the $y=0$ axis also exists but is not plotted), while escape from the central region happens when the evolved orbit crosses the black colored arcs of circles centered at the origin and having  as radius the distance of the outermost point of the Lyapunov orbit from the center.
	}
	\label{fig:open}
\end{figure}

Since the properties of the openings in the vicinity of the $L_1$ and $L_2$ points influence the dynamics of escapes from (or entries to) the central region, we can formulate a criterion for determining such behaviors based on the features of the Lyapunov orbits near the $L_1$ and $L_2$ points. We note that  these orbits have already been used to define whether an orbit has escaped the central region \cite{C90,KLMR01a,KLMR01b}. In Fig.~\ref{fig:open}  we plot, in the configuration space of system \eqref{eq:2DHam} with $\Omega_{\mathrm{b}}=60\,\mbox{km} \,\mbox{s}^{-1}\, \mbox{kpc}^{-1}$, the Lyapunov periodic orbit (blue curves) located at the opening between the energetically forbidden regions (gray colored areas) with $y>0$ (the symmetric with respect to axis $y=0$ Lyapunov orbit is not depicted) for energy values  $E=-4.15$ [Fig.~\ref{fig:open}(a)], $E=-4.2$ [Fig.~\ref{fig:open}(b)] and $E=-4.235$ [Fig.~\ref{fig:open}(c)]. Since an orbit starting at the  central region escapes to the exterior region once its distance from the center becomes larger than the distance  $D$ of the utmost point of the Lyapunov periodic orbit located at the opening \cite{KLMR01a}, we consider that an orbit escapes when it crosses, moving outwards, a circle of radius $D$ centered at $(x,y)=(0,0)$. As such escapes can happen only through the openings between the forbidden regions, we plot in Fig.~\ref{fig:open} only the arcs of these circles (black curves between the gray colored regions). In each panel of Fig.~\ref{fig:open} we also depict the time evolution of an orbit starting at point $(x,y)=(0.88, 2.48)$ (indicated by an orange dot), with initial velocities $\dot{y}=0$ and $\dot{x}\geq 0$ calculated from \eqref{eq:2DHam},  until a short time after  its escape. We see that for the lower energy $E=-4.235$ [Fig.~\ref{fig:open}(c)],  for which the size of the openings is small, the orbit spends a lot of time at the central region before escaping from the opening with $y<0$. As the energy increases to $E=-4.2$ [Fig.~\ref{fig:open}(b)] and $E=-4.15$ [Fig.~\ref{fig:open}(a)], the size of the openings grows as the area of the forbidden regions diminishes, and, in general, it becomes easier for orbits to escape faster.

We note that for the numerical integration of the orbits presented in Fig.~\ref{fig:open}, as well as all for all the other orbits computed  in our work, we used  a sixth order Runge-Kutta method \cite{L68}, with  integration time steps $dt \approx 10^{-3}$, which allowed us to keep the relative energy error below $10^{-9}$. 

\section{Lagrangian descriptors and invariant manifolds}
\label{sec:LDs}

Let us consider an orbit of the Hamiltonian system \eqref{eq:2DHam} with initial condition $\vec{x}_0=(x(0), y(0), \dot{x}(0), \dot{y}(0))$ at time $t=0$, whose evolution is governed by the related Hamilton equations of motion 
\begin{equation}
	\dot{\vec{x}} = \frac{d \vec{x}} {d t} = \vec{f}(\vec{x}, t). 
	\label{eq:eqmotion}
\end{equation}
Then, following \cite{LBGWM17} we define the orbit's LD as 
\begin{equation}
	M(\vec{x}_0,\tau)=\int_{-\tau}^{\tau}\sum_{i=1}^{4}\big|f_i(\vec{x},t)\big|^p dt,
	\label{eq:LDdef}
\end{equation}
where $f_i(\vec{x},t)$ is the $i^{\text{th}}$ component of the vector field $\vec{f}(\vec{x},t)$ \eqref{eq:eqmotion}, and $\tau$ is the length of the integration time interval, both forward and backward in time. In our work we set $p=0.5$ as this value  has been successfully implemented in various studies (e.g.,~\cite{DW17,KGAW20}), allowing the efficient identification  of manifolds, as well as pronouncing discontinuities of the gradient of the LD values in the systems' phase space. We note that the evaluation of the LD \eqref{eq:LDdef} can be separated into its  forward
\begin{equation}
	M^{\mathrm{f}}(\vec{x}_0,\tau)=\int_{0}^{\tau}\sum_{i=1}^{4}\big|f_i(\vec{x},t)\big|^p dt,
	\label{eq:LDforward}
\end{equation}
and backward 
\begin{equation}
	M^{\mathrm{b}}(\vec{x}_0,\tau)=\int_{-\tau}^{0}\sum_{i=1}^{4}\big|f_i(\vec{x},t)\big|^p dt,
	\label{eq:LDbackward}
\end{equation}
in time components, so that 
\begin{equation}
	M(\vec{x}_0,\tau)=M^{\mathrm{f}}(\vec{x}_0,\tau) +  M^{\mathrm{b}}(\vec{x}_0,\tau).
	\label{eq:LDsum}
\end{equation}

\begin{figure}[tb!]
	\centering
		\includegraphics[width=\textwidth]{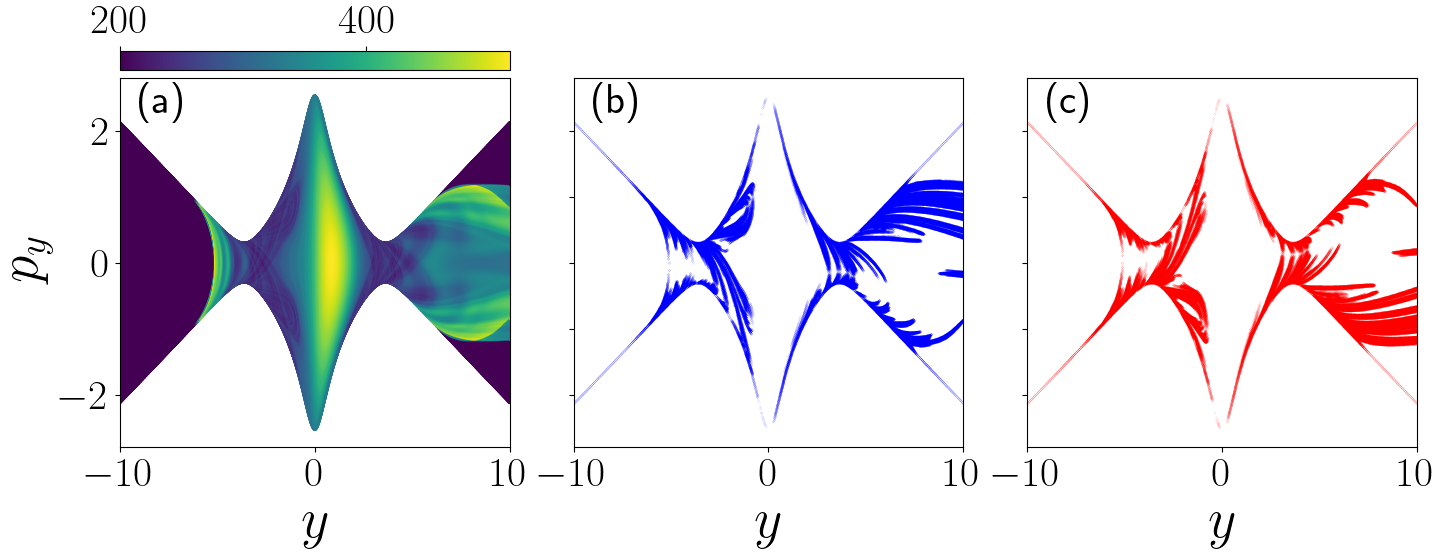}    
	\caption{Results obtained for orbits having their initial conditions on a $2000\times2000$ grid in the region $y\in[-10,10]$ and $p_y\in[-2.8,2.8]$ of the $x=0$, $p_x\geq 0$ PSS of system \eqref{eq:2DHam} with $H=-4.2$ and $\Omega_{\mathrm{b}}=60\,\mbox{km} \,\mbox{s}^{-1}\, \mbox{kpc}^{-1}$. In (a) each initial condition is colored according to the orbit's LD \eqref{eq:LDdef} for $\tau=60 \, t_{\mathrm{u}}$, following the color scale at the top of the panel. White colored areas denote energetically forbidden regions. The set of stable [unstable] manifolds extracted from the scalar field of forward [backward] LD $M^{\mathrm{f}}$  \eqref{eq:LDforward} [$M^{\mathrm{b}}$ \eqref{eq:LDbackward}] as the grid points having a LD gradient larger than $d=0.8$, is shown in (b) [(c)].}
	\label{fig:LDs_manifolds}
\end{figure}

In order to visualize the phase space structure of system \eqref{eq:2DHam} with $E=-4.2$ and $\Omega_{\mathrm{b}}=60\,\mbox{km} \,\mbox{s}^{-1}\, \mbox{kpc}^{-1}$, we compute the LD values for a dense set of initial conditions on the system's Poincar\'{e} surface of section (PSS) defined by $x=0$, $p_x= \dot{x} - \Omega_{\mathrm{b}} y \geq 0$, through \eqref{eq:LDdef} for $\tau=60 \, t_{\mathrm{u}}$, and color each on of them according to its LD value. The outcome of this process is shown in  Fig.~\ref{fig:LDs_manifolds}(a), which is constructed by considering a grid of $2000\times2000$ equally spaced initial conditions in the region defined by $-10 \leq y \leq 10$ and  $-2.8 \leq p_y \leq 2.8$, where $p_y = \dot{y} + \Omega_{\mathrm{b}} x$.  In this figure we see an alternative representation to the one depicted in Fig.~\ref{fig:open}(b), of the various regions where motion is permitted. The $x=0$ part of the central region of Fig.~\ref{fig:open}(b) corresponds to the rhomboidal area in the middle of Fig.~\ref{fig:LDs_manifolds}(a). This region is connected to the exterior region, represented by the two triangular looking areas at the right and the left edges of Fig.~\ref{fig:LDs_manifolds}(a), through two narrow  openings related to the $L_1$ ($y>0$) and $L_2$ ($y<0$) points. Although the shape of the permitted region of motion is symmetric with respect to the $y=0$ axis,  the structures revealed through the LD computations do not preserve this symmetry as the considered orbits have a specific initial orientation defined by $p_x \geq 0$.

In Fig.~\ref{fig:LDs_manifolds}(a) we can identify phase space regions characterized by high or low LD values, while stable and unstable invariant manifolds  appear as `singular features' in the LDs'   scalar field where the descriptor values change abruptly. The location of these manifolds can become apparent by selecting the grid points where the (positive defined)  gradient of the LDs field is larger than a threshold value $d$. Applying this approach for $d=0.8$   to the results of Fig.~\ref{fig:LDs_manifolds}(a) which are obtained from the forward time LDs $M^{\mathrm{f}}$ \eqref{eq:LDforward} we present the set of the system's stable manifolds in Fig.~\ref{fig:LDs_manifolds}(b), while the unstable manifolds, computed through the evaluation of the  backward time LDs $M^{\mathrm{b}}$ \eqref{eq:LDbackward}, are shown in Fig.~\ref{fig:LDs_manifolds}(c). 

\section{The effect of invariant manifolds on the system's orbital behavior}
\label{sec:Total_effect}

In this section we investigate, following several diverse approaches, the effect of the system's stable and unstable manifolds, whose structure and location are revealed through LD computations similar to the ones presented in  Fig.~\ref{fig:LDs_manifolds}, on the orbital dynamics of Hamiltonian \eqref{eq:2DHam}.

\subsection{Features of escape dynamics}
\label{sec:escape}

We first investigate the influence of the system's manifolds on the way that orbits starting in the central region of the configuration space might escape to the exterior region. In particular, we study the time $t_{\mathrm{e}}$ required for an orbit to escape from the central region, compute the duration $t_{\mathrm{d}}$ that this orbit might remain outside the central region within a given total time interval, as well as register the number $N_{\mathrm{e}}$ of times that an orbit escapes from the central region. In general, we can follow the evolution of orbits initially located at the system's central region both forward or backward in time, and study the features of escapes for both evolutions. Nevertheless, due to the symmetry between of the stable and unstable manifolds on the system's PSS with respect to the $p_y=0$ axis [the unstable manifolds shown in Fig.~\ref{fig:LDs_manifolds}(c) can be obtained by reflecting the stable manifolds of Fig.~\ref{fig:LDs_manifolds}(b) about the $p_y=0$ axis], we consider only  the forward in time evolution of orbits. Since, as was also mentioned in \cite{RMAG06}, the stable manifolds influence the forward in time evolution of orbits from the central region toward the Lyapunov orbits, we investigate the connection between the features of escape dynamics only with the morphology of the system's stable manifolds.

\begin{figure}[tb!]
	\centering
		\includegraphics[width=\textwidth]{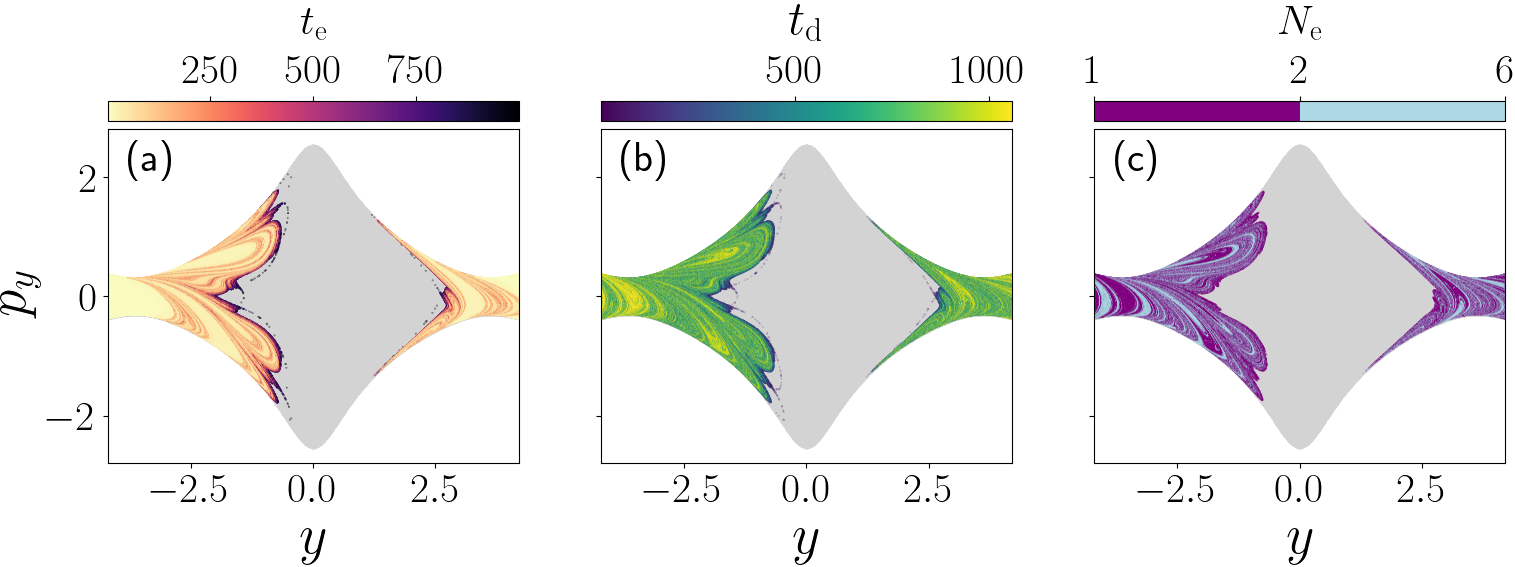}    
		\includegraphics[width=\textwidth]{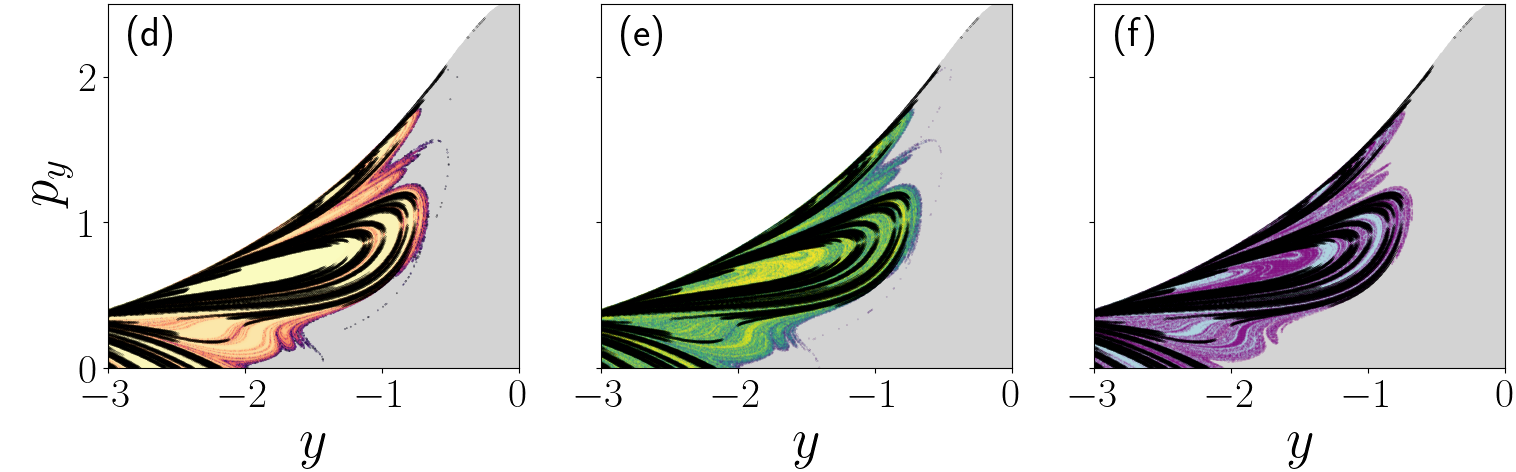}    
	\caption{Initial conditions on the $x=0$, $p_x\geq 0$ PSS of system \eqref{eq:2DHam} with $E=-4.2$ and $\Omega_{\mathrm{b}}=60\,\mbox{km} \,\mbox{s}^{-1}\, \mbox{kpc}^{-1}$, colored according to [(a) and (d)] the time $t_{\mathrm{e}}$ needed for the related orbits' first escape to the external  region of the configuration space, [(b) and (e)] the total time duration $t_{\mathrm{d}}$ the  orbits spends at the exterior region in the first $t=1071 \, t_{\mathrm{u}}$ ($\approx 5 \mbox{Gyrs}$) of the evolution, [(c) and (f)]  the number $N_{\mathrm{e}}$ of escapes (transitions from the central to the exterior region) experienced by the orbits until $t=1071 \, t_{\mathrm{u}}$. Panels (a), (b) and (c) are created using initial conditions on a $2000\times2000$ grid in the region $y\in[-4.2,4.2]$ and $p_y\in[-2.8,2.8]$, while panels (d), (e) and (f) present a zoom-in view, of respectively panels (a), (b) and (c), in the region defined by  $-3 \leq y \leq 0$ and $ 0 \leq p_y \leq 2.5$. Points in panels belonging to the same column of plots are colored according to  the color scale at the top of the upper panel. Note that in panels (c) and (f) we use only two colors: purple when $N_{\mathrm{e}} \leq 2$, and cyan when $2 <N_{\mathrm{e}} \leq 6$.  Initial conditions corresponding  to orbits which do not escape over the total integration time are colored in gray, while white colored areas denote energetically forbidden regions. In (d), (e) and (f) the stable manifolds of Fig.~\ref{fig:LDs_manifolds}(b) are plotted in black. 
	}
	\label{fig:escapes}
\end{figure}

In Fig.~\ref{fig:escapes}(a) we color (using the color scale at the top of the figure) initial conditions in the $x=0$, $p_x\geq 0$ PSS of system \eqref{eq:2DHam} with $E=-4.2$ and $\Omega_{\mathrm{b}}=60\,\mbox{km} \,\mbox{s}^{-1}\, \mbox{kpc}^{-1}$, according to the time $t_{\mathrm{e}}$ needed for the corresponding orbit to escape to the exterior region in the system's configuration space. A zoom-in of a part of the PSS presented in Fig.~\ref{fig:escapes}(a), corresponding to $y<0$, $p_y>0$, is shown in Fig.~\ref{fig:escapes}(d), where also the stable manifolds computed in Fig.~\ref{fig:LDs_manifolds}(b) are plotted in black. We note  that all considered initial conditions in Figs.~\ref{fig:escapes}(a) and (d), as well as in all panels of Fig.~\ref{fig:escapes}, are integrated up to a final time $t=1071 \, t_{\mathrm{u}}$, which approximately corresponds to $ 5 \mbox{Gyrs}$. Furthermore, in Fig.~\ref{fig:escapes} energetically forbidden phase space regions are colored in white, while gray colored initial conditions correspond to orbits that do not escape to the exterior region over the considered integration time interval.

In Figs.~\ref{fig:escapes}(a) and (d) we observe the existence of filament-like, lobe  forming structures inside the regions where escapes occur. In these figures initial conditions which lead to fast escapes, characterized by small $t_{\mathrm{e}}$ values, are colored in light yellow. Although such regions appear to mainly occupy areas close to the escape openings, they also exist in  regions with smaller $|y|$ values, indicating that also orbits  closer to the central region can escape fast, due to the influence of the stable manifolds, which can guide orbits toward the exterior regions. As  becomes clear from  Fig.~\ref{fig:escapes}(d), the morphology of the stable manifolds exhibit similarities with the structure of the color plot created by the orbits' $t_{\mathrm{e}}$ values, indicating that these manifolds are shaping the escape orbital behavior of the system. It is worth noting that we do not observe in Fig.~\ref{fig:escapes}(d) a one-to-one correspondence between the manifolds and the features of the underlying color map, although the overall behavior of the two structures are similar. This small discrepancy is due to the fact that the depicted manifolds are obtained through a rather short time integration ($\tau=60 \, t_{\mathrm{u}}$) of orbits, while the escape dynamics picture created through the computation of $t_{\mathrm{e}}$ values was obtained for much longer integration times (up to $t=1071 \, t_{\mathrm{u}}$).

The stable manifolds not only influence the time $t_{\mathrm{e}}$ of the first escape of an orbit from the central to the exterior regions of the system's configuration space, but also affect other features of the orbits' evolution. An orbit that escapes to the exterior region can also reenter the central region, by passing through  the openings near the $L_1$ and $L_2$ points. Actually, there exist orbits which exit and reenter the central region many times. In Figs.~\ref{fig:escapes}(b) and (e) we color initial conditions according to the total time $t_{\mathrm{d}}$ that an orbit spends at the exterior region up until the final integration time of our simulations ($t=1071 \, t_{\mathrm{u}}$). In these figures we again observe, as in Figs.~\ref{fig:escapes}(a) and (d), the appearance of filament, lobe creating structures, which, once more, are in good correlation with the form of the stable manifolds [Figs.~\ref{fig:escapes}(e)].

Furthermore,  in Figs.~\ref{fig:escapes}(c) and (f)  we color initial conditions according to the number $N_{\mathrm{e}}$ of escapes from the central region that the corresponding orbits experience during their evolution. Since $N_{\mathrm{e}}$ gets discrete values we prefer to divide the initial conditions in two  main categories. Thus, we have orbits which exhibit a small number of escapes, namely $N_{\mathrm{e}} \leq 2$, which are colored in purple.  In principle, these orbits spend a large part of their evolution in the exterior region and consequently should have large $t_{\mathrm{d}}$ values. This relation between $N_{\mathrm{e}}$ and $t_{\mathrm{d}}$  is also reflected to  the good correspondence between the regions colored in  purple in Figs.~\ref{fig:escapes}(c) and (f) and in light yellow (large  $t_{\mathrm{d}}$  values) in Figs.~\ref{fig:escapes}(b) and (e). Orbits exhibiting $N_{\mathrm{e}} > 2$ number of escapes from the central region (and consequently a large number of reentries into  it) are colored in cyan in Figs.~\ref{fig:escapes}(c) and (f).  Again the structure of the color plots of the $N_{\mathrm{e}}$  values  is in good agreement with the morphology of the stable manifolds which are plotted in black in Fig.~\ref{fig:escapes}(f).

\subsection{The past and future evolution of orbits}
\label{sec:past-future_1}

\begin{figure}[tb!]
	\centering
		\includegraphics[width=\textwidth]{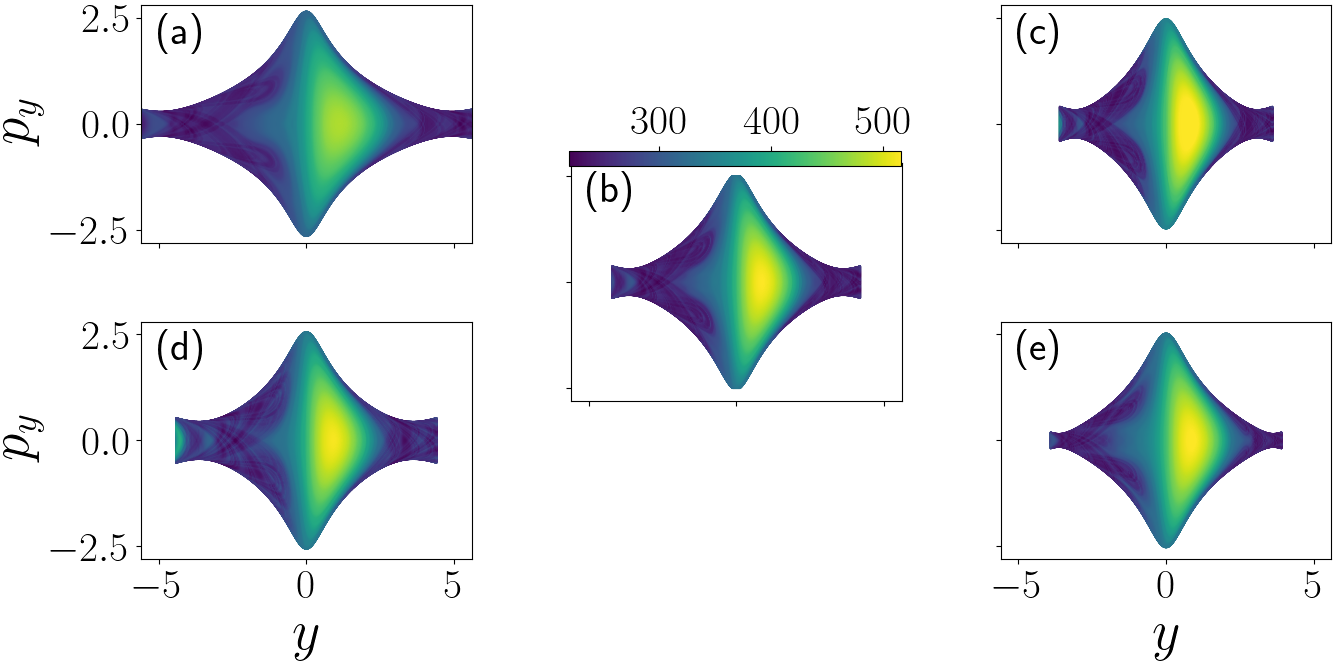}    
	\caption{Initial conditions on the $x=0$, $p_x\geq 0$ PSS of system \eqref{eq:2DHam}  colored according to the related orbit's LD \eqref{eq:LDdef} for $\tau=60 \, t_{\mathrm{u}}$, following the color scale at the top of the panel (b). In each panel a grid of $2000 \times  2000$ equally spaced initial conditions were considered in the region $-2.8 \leq p_y \leq  2.8$, with the $y$ range being adjusted to cover the distance between the outermost points off the two Lyapunov orbits. The considered energy $E$ and pattern speed $\Omega_{\mathrm{b}}$  values are: (a) $E=-3.91$ and $\Omega_{\mathrm{b}}=45\,\mbox{km} \,\mbox{s}^{-1}\, \mbox{kpc}^{-1}$, (b) $E=-4.2$ and $\Omega_{\mathrm{b}}=60\,\mbox{km} \,\mbox{s}^{-1}\, \mbox{kpc}^{-1}$, (c) $E=-4.37$ and $\Omega_{\mathrm{b}}=70\,\mbox{km} \,\mbox{s}^{-1}\, \mbox{kpc}^{-1}$, (d) $E=-4.15$ and $\Omega_{\mathrm{b}}=60\,\mbox{km} \,\mbox{s}^{-1}\, \mbox{kpc}^{-1}$, and (e) $E=-4.235$  and $\Omega_{\mathrm{b}}=60\,\mbox{km} \,\mbox{s}^{-1}\, \mbox{kpc}^{-1}$. These parameters were selected such that the size of openings in (a), (b) and (c) are practically equal,  while panels (d), (b) and (e) respectively correspond to Figs.~\ref{fig:open}(a), (b) and (c).}
	\label{fig:5LDplots}
\end{figure}

In order to investigate the effect of different parameters of Hamiltonian \eqref{eq:2DHam}, like the system's energy $E$ and pattern speed $\Omega_{\mathrm{b}}$, on the structure of the invariant manifolds, and  consequently on the model's orbital behavior, we present in Fig.~\ref{fig:5LDplots} color plots based on the values of LDs for $\tau=60 \, t_{\mathrm{u}}$. The size of the openings through which orbits can escape from the central region (or reenter to it) depends both on the system's energy and  pattern speed. To study the effect of the pattern speed on the morphology of the stable and unstable manifolds we consider in Figs.~\ref{fig:5LDplots}(a), (b) and (c) cases with different $\Omega_{\mathrm{b}}$  and appropriately chosen energy values such that the size of the openings  are practically the same in all cases. In particular, consider arrangements with $E=-3.91$ and $\Omega_{\mathrm{b}}=45\,\mbox{km} \,\mbox{s}^{-1}\, \mbox{kpc}^{-1}$ [Fig.~\ref{fig:5LDplots}(a)], $E=-4.2$ and $\Omega_{\mathrm{b}}=60\,\mbox{km} \,\mbox{s}^{-1}\, \mbox{kpc}^{-1}$ [Fig.~\ref{fig:5LDplots}(b)], and $E=-4.37$ and $\Omega_{\mathrm{b}}=70\,\mbox{km} \,\mbox{s}^{-1}\, \mbox{kpc}^{-1}$ [Fig.~\ref{fig:5LDplots}(c)]. We see that as the pattern speed is increased the system's energy is decreased in order to achieve openings of equal size. From the comparison of Figs.~\ref{fig:5LDplots}(a), (b) and (c) (mainly of the regions with $y<0$) we observe that a rise  in $\Omega_{\mathrm{b}}$ (decrease in $E$) leads to an increase in the intricacy of the manifolds. Furthermore, keeping the pattern speed constant to $\Omega_{\mathrm{b}}=60\,\mbox{km} \,\mbox{s}^{-1}\, \mbox{kpc}^{-1}$ and decreasing the system's energy from $E=-4.15$ [Fig.~\ref{fig:5LDplots}(d)] to $E=-4.2$ [Fig.~\ref{fig:5LDplots}(b)] and eventually to $E=-4.235$ [Fig.~\ref{fig:5LDplots}(e)] [i.e.,  considering the cases respectively presented in Figs.~\ref{fig:open}(a), (b) and (c)] leads to a decrease of the openings' size, along with a decline in the manifolds' complexity.

\begin{figure}[tb!]
	\centering
		\includegraphics[width=\textwidth]{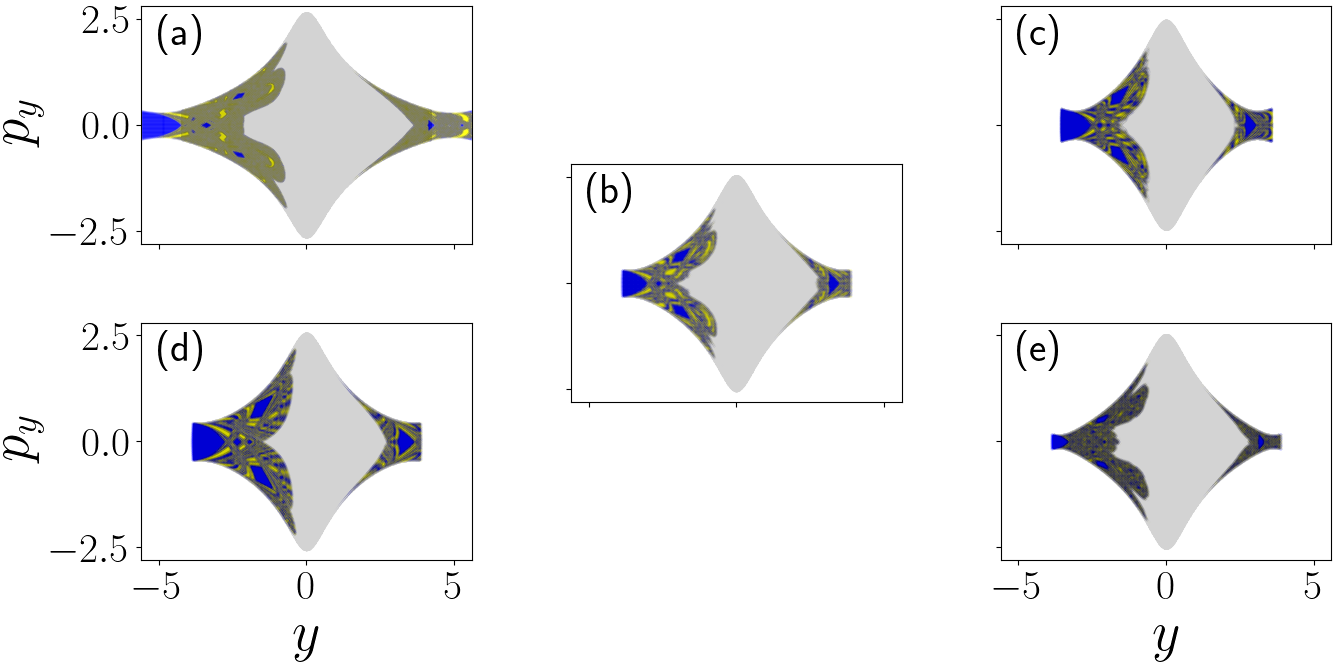}    
	\caption{Similar to Fig.~\ref{fig:5LDplots}, with initial conditions colored in blue (yellow) if the corresponding orbits entered in the past to the central region and will escape from it in the future, from the same (different) opening  between the energetically forbidden regions [i.e., orbits being homoclinic (heteroclinic)], when orbits are integrated both backward and  forward in time for $t=289 \, t_{\mathrm{u}}$ ($\approx 1.35 \mbox{Gyrs}$).  The percentage $P_{\mathrm{hom}}$ of homoclinic orbits is (a) $P_{\mathrm{hom}} = 50.92\%$, (b) $P_{\mathrm{hom}} = 52.03\%$, (c) $P_{\mathrm{hom}} = 56.58\%$, (d) $P_{\mathrm{hom}} = 55.97\%$, and (e) $P_{\mathrm{hom}} = 52.94\%$.  }
	\label{fig:5homhet}
\end{figure}

In order to investigate how  the history of an orbit influences its future evolution, we integrate several initial conditions, both forward and backward in time, and analyze their behaviors. This approach is also dictated by our desire  to understand the effect of the system's stable and unstable manifolds (which can be extracted through respectively the forward and backward in time computation of LDs) on the model's orbital characteristics. In \cite{RAMG07} orbits of a galactic potential were called  \textit{homoclinic} (\textit{heteroclinic}) if during their evolution they were escaping from the central region of the system through a particular opening between the energetically forbidden areas, and later on they were reentering the central region from the same (opposite) opening. In that work the importance of different types of orbits, namely homoclinic, heteroclinic and simply escaping orbits, for supporting different morphological features of the galaxy, like ring structures and spiral arms,  was emphasized. Inspired by these ideas we introduce a slightly different definition of the terms homoclinic heteroclinic in our study.  In particular, we name an orbit homoclinic (heteroclinic) if it escapes from the central region from the same (different) opening when it is integrated both forward and backward in time. We perform this analysis for the orbits and the cases considered in Fig.~\ref{fig:5LDplots} using a final (both backward and forward) integration time $t=289 \, t_{\mathrm{u}}$ ($\approx 1.35 \mbox{Gyrs}$), or until the first escape from the cental region occurs. We choose a smaller time interval than the one considered in Fig.~\ref{fig:escapes} (i.e.~$t=1071 \, t_{\mathrm{u}}$), to not only  reduced the computational cost, but mainly because it allows us to reveal  some well pronounced structures in the system's configuration space. The outcomes of this process are presented in Fig.~\ref{fig:5homhet}, whose panels correspond to the same cases shown in Fig.~\ref{fig:5LDplots}.  In each panel of Fig.~\ref{fig:5homhet} initial conditions leading to homoclinic (heteroclinic) behavior are colored in blue (yellow), while, as usual, orbits which do not escape during the simulation are colored in gray.

The results of Figs.~\ref{fig:5homhet}(a), (b) and (c) indicate that as the pattern speed $\Omega_{\mathrm{b}}$ increases, with the system's energy $E$ decreasing to keep the openings' size practically constant, the blue and yellow colored regions, respectively corresponding to homoclinic and heteroclinic escaping orbits, become less mixed. Consequently, connected phase space regions leading to similar dynamical behaviors increase in size. In addition, the percentage of escaping homoclinic orbits $P_{\mathrm{hom}}$ grows from $P_{\mathrm{hom}} = 50.92\%$ [Fig.~\ref{fig:5homhet}(a)], to $P_{\mathrm{hom}} = 52.03\%$ [Fig.~\ref{fig:5homhet}(b)], and eventually to $P_{\mathrm{hom}} = 56.58\%$ [Fig.~\ref{fig:5homhet}(c)], indicating that as more connected and larger phase space regions display similar dynamical trends more orbits enter the central region and escape from it passing from the same opening between the energetically forbidden areas. 

The comparison of Figs.~\ref{fig:5homhet}(d), (b) and (e) show that the decrease of the system's energy $E$ for a fixed pattern speed $\Omega_{\mathrm{b}}=60\,\mbox{km} \,\mbox{s}^{-1}\, \mbox{kpc}^{-1}$ leads to the higher mixing of the differently colored points, or, in other words, an increased sensitivity of the orbital evolution on the location of the initial condition. In addition, this energy decrease is  followed by a change of the percentage of homoclinic orbits from $P_{\mathrm{hom}} = 55.97\%$ [Fig.~\ref{fig:5homhet}(d)], to $P_{\mathrm{hom}} = 52.03\%$ [Fig.~\ref{fig:5homhet}(b)], and  $P_{\mathrm{hom}} = 52.94\%$ [Fig.~\ref{fig:5homhet}(e)]. 

The direct comparison of Figs.~\ref{fig:5LDplots} and   \ref{fig:5homhet}, suggest that the increase of the complexity of the manifold structures [i.e., moving from the case presented in Figs.~\ref{fig:5LDplots}(a) and   \ref{fig:5homhet}(a) to the one shown in Figs.~\ref{fig:5LDplots}(c) and   \ref{fig:5homhet}(c), as well as transitioning from the arrangement of Figs.~\ref{fig:5LDplots}(e) and   \ref{fig:5homhet}(e) to the case of Figs.~\ref{fig:5LDplots}(d) and   \ref{fig:5homhet}(d)]   results in the creation of more well defined boundaries between phase space regions characterized by homoclinic or heteroclinic  behaviors, and in the diminishing of the sensitivity of orbital behaviors on the location of  initial conditions, as the size of connected phase space areas having the same dynamical behavior  increases. It is worth noting that, in general, these trends are also followed by an increase of $P_{\mathrm{hom}}$.

\subsubsection{Study of the past and future orbital dynamics via the origin-fate map}
\label{sec:past-future_2}

Trying to understand in depth the distributions of homoclinic and heterocyclic orbits in Fig.~\ref{fig:5homhet}, we further take into account the specific opening from which an orbit escapes from the central region  (determined through the orbit's forward time evolution), as well as, the opening through which it entered the central region (obtained through the backward in time integration of the orbit). In particular, we attribute to the opening related with the $L_1$ point the  index $j=1$,  while the other opening, appearing in the vicinity of the $L_2$ point, is represented by $j=2$. Then, for  orbits experiencing escapes from the central region in both the forward and backward time  evolutions we keep track of the opening between the energetically forbidden areas from which they escape in forward time (called the orbit's fate), as well as in backward time (referred to as  the orbit's origin), and  associate two indices to each initial condition leading to escapes. The first index indicates the orbit's origin, while the second denotes its fate. Coloring all possible combinations of indices with different colors, we obtain a representation of the system's dynamics by creating its OFM. The notion of the OFM  was first introduced in \cite{HKWS23} where the phase space transport in a reactant-product type chemical dynamical system was studied. 

\begin{figure}[tb!]
	\centering
		\includegraphics[width=\textwidth]{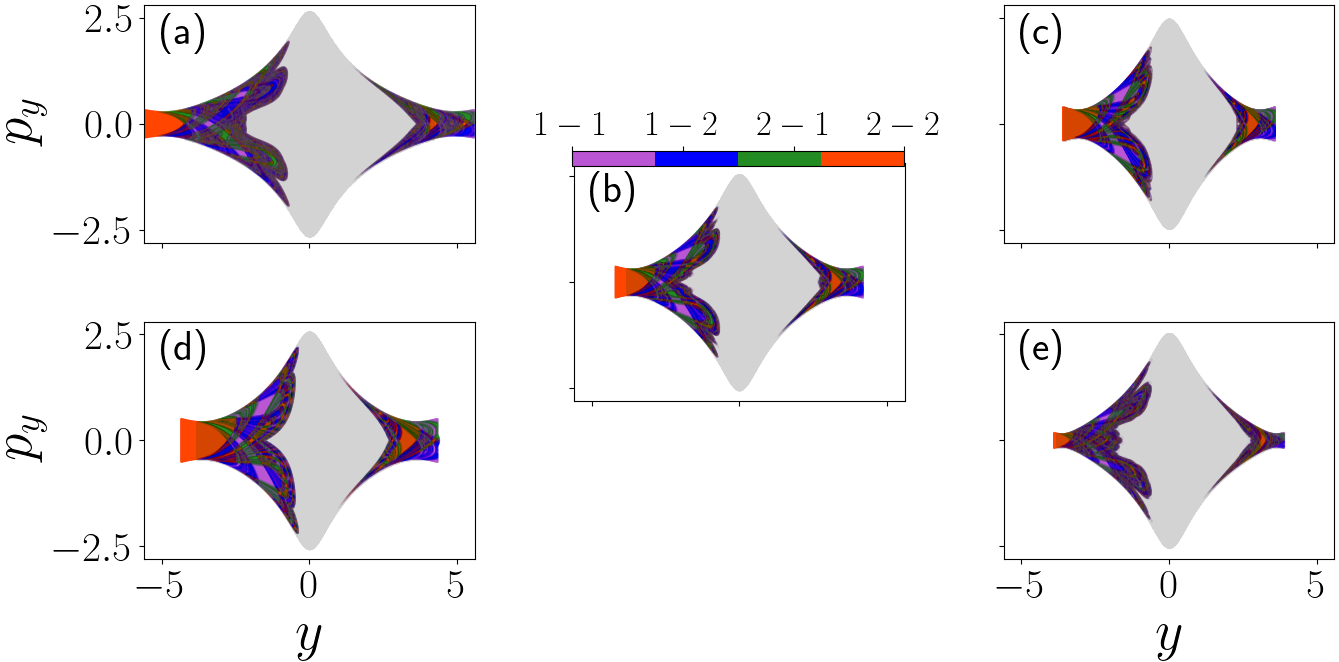}    
	\caption{The OFMs of the cases shown in Fig.~\ref{fig:5homhet} where initial conditions are colored according to their origin-fate indices $j$, following the color arrangements indicated above (b). In these arrangements the first index indicates the opening from which the orbit entered the central region (or equivalently escaped from it when it was integrated backward in time), while the second index denotes the opening through which the orbit escapes to the external region when it is integrated forward in time. Index $j=1$ corresponds to the opening close to the $L_1$ point, while $j=2$ denotes the opening close to the $L_2$ point. }
	\label{fig:5OFM}
\end{figure}

The OFMs for the cases presented in Fig.~\ref{fig:5homhet} are given in Fig.~\ref{fig:5OFM}, where for example initial conditions leading to orbits which entered the central region in the past from opening $j=1$ and  eventually depart from it through the opening with index $j=2$ (corresponding to the combination 1-2) are colored in blue. The other possible arrangements correspond to points colored in purple (1-1), green (2-1) and red  (2-2). The plots of Fig.~\ref{fig:5OFM} add  additional information on their counterparts in Fig.~\ref{fig:5homhet} about the specific routes the orbits follow to enter and exit the central region. For example, we see that as the pattern speed $\Omega_{\mathrm{b}}$ is increased moving from Fig.~\ref{fig:5OFM}(a) to Fig.~\ref{fig:5OFM}(b), and eventually to Fig.~\ref{fig:5OFM}(c), regions exhibiting the same dynamical behavior become more connected and increase their size, while the area covered by scattered, mixed colored points decreases. Furthermore, the comparison of Figs.~\ref{fig:5OFM}(d) and \ref{fig:5OFM}(e) indicate that, as the opening between the energetically forbidden region decreases in size, while $\Omega_{\mathrm{b}}$ is kept constant, the sensitivity of the observed orbital behavior on the location of the orbit's initial condition grows.

\begin{figure}[tb!]
	\centering
		\includegraphics[width=\textwidth]{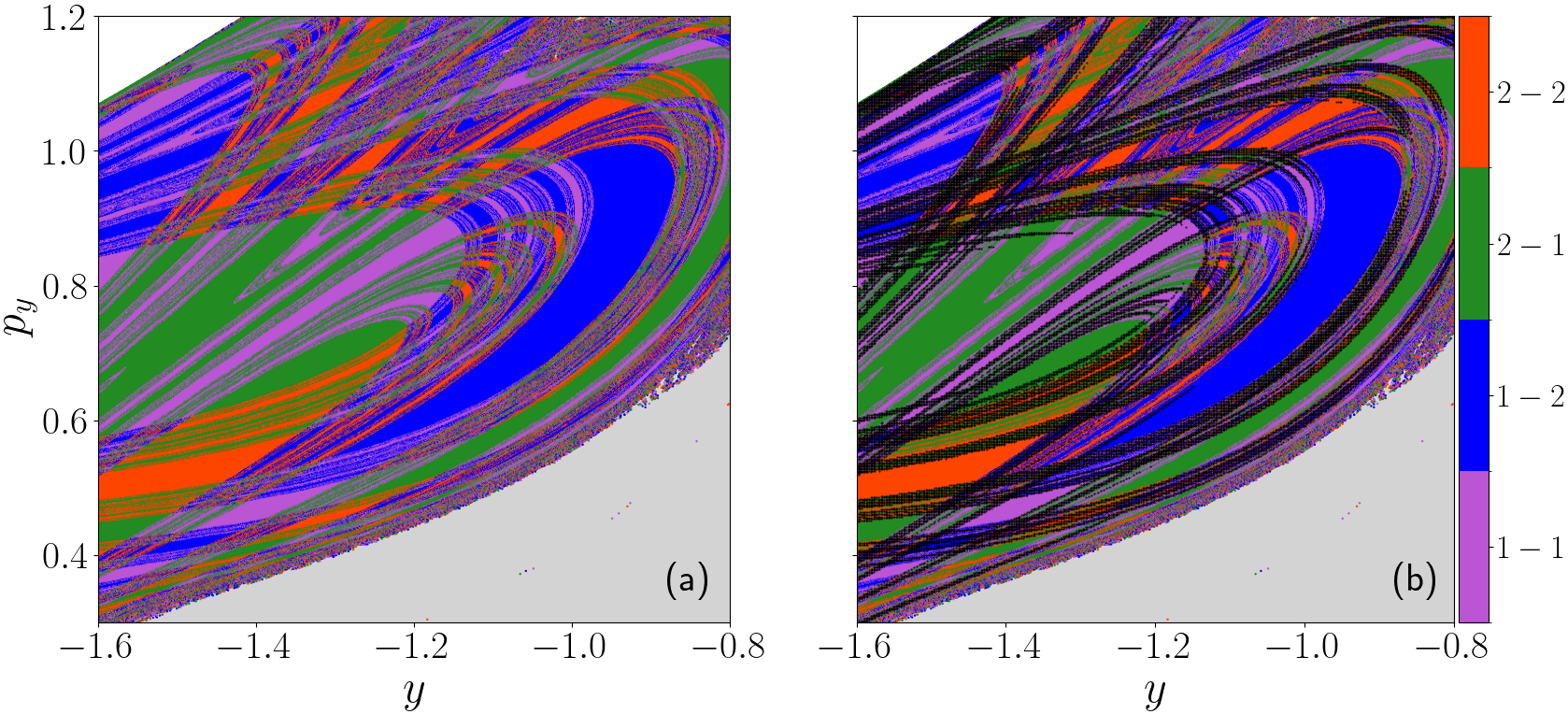}    
	\caption{(a) A portion of the OFM presented in Fig.~\ref{fig:5OFM}(b)  defined by $y \in [-1.6,  -0.8]$ and $ p_y \in [0.3,  1.2]$. (b) Similar to (a) along with the addition of the stable and unstable  manifolds  computed through the evolution of the orbits' LDs following the same process used in panels of Figs.~\ref{fig:LDs_manifolds} and \ref{fig:escapes}. In both panels  points are colored according to the color bar on the right side of (b).}
	\label{fig:OFMzoom}
\end{figure}

In order to study the influence of the manifolds' structure  on the shape of the OFM, as well as  the dynamical behavior of orbits, we focus our attention to a portion of the phase space depicted in Fig.~\ref{fig:5OFM}(b), which is  defined by $-1.6 \leq y \leq -0.8$ and $0.3 \leq p_y \leq  1.2$. The OFM  for this region, calculated on a grid of $2000 \times 2000$ evenly spaced initial conditions, when orbits are integrated for $t=289 \, t_{\mathrm{u}}$  forward and backward in time, is presented in Fig.~\ref{fig:OFMzoom}(a). The relation of the complicated structures of the various colored  regions shown in Fig.~\ref{fig:OFMzoom}(a) with the features of the system's stable and unstable manifolds  becomes apparent in Fig.~\ref{fig:OFMzoom}(b) where we superimpose these manifolds to the OFM by plotting them in black. These manifolds are obtained by following exactly the same procedure we used in order to construct the manifolds presented in various panels of Figs.~\ref{fig:LDs_manifolds} and \ref{fig:escapes}. We note that some structures observed in the color map of  Fig.~\ref{fig:OFMzoom}(a) are not outlined in Fig.~\ref{fig:OFMzoom}(b) by the computed manifolds, because these manifolds were evaluated for a small time interval, namely $\tau=60 \, t_{\mathrm{u}}$, while the OFM was constructed by using much longer time evolutions of orbits ($t=289 \, t_{\mathrm{u}}$ both forward and backward in time). An increase in the time considered for the computation of the system's LDs, will lead to the revelation of more structural details of the manifolds, but simultaneously will create a very entangled picture of curves, obscuring, in some sense, the manifolds features. For this reason we did not perform such computations. 

\subsection{Morphological features in the system's configuration space}
\label{sec:morphological_struct}

The importance of the influence of manifolds in dictating the dynamics of  orbits which support ring and spiral structures in galaxies has been emphasized in several papers (see for example \cite{RMAG06,RAMG07,ARM09,ARBM09}). Motivated by such studies we  explore in this section the possible relation of orbits from different phase space regions to the formation of specific galactic morphological features, as well as the dependence of these features on the system's energy $E$, pattern speed $\Omega_{\mathrm{b}}$, as well as the size of the openings between the energetically forbidden regions.

In Sect.~\ref{sec:escape} we investigated the relation between the structure of stable and unstable manifolds to different aspects of the orbital behavior of ensembles of initial conditions, namely the time $t_{\mathrm{e}}$ an orbit needs to escape from the central region, the total time duration $t_{\mathrm{d}}$ an orbit spends in the exterior region, as well as the number $N_{\mathrm{e}}$ of escapes from the central region an orbit experiences in some given time interval. From the results of Fig.~\ref{fig:escapes} we saw that all these quantities are, more or less, similarly influenced by the structure of the system's manifolds. Thus, we now focus  only on investigating the possible connection between $N_{\mathrm{e}}$ and the morphological features appearing in the system's configuration space.

\begin{figure}[tb!]
	\centering
		\includegraphics[width=\textwidth]{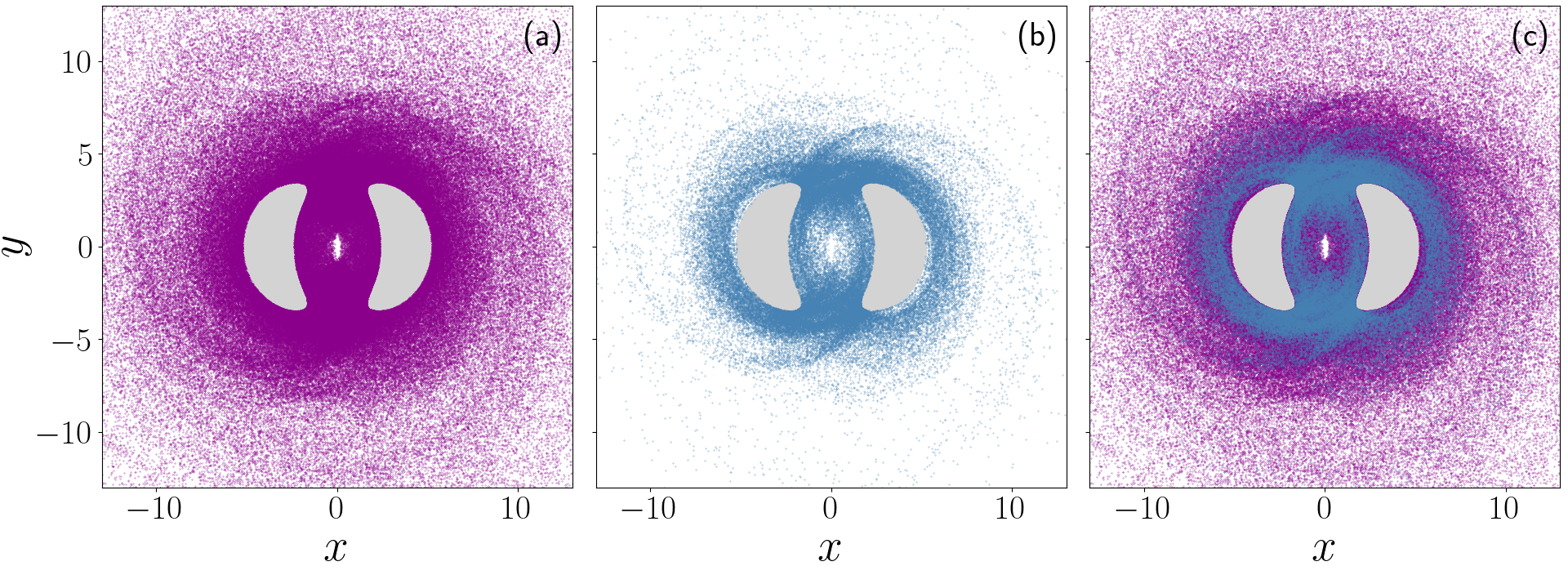}    
	\caption{Final positions of stars in system's \eqref{eq:2DHam} configuration space $(x,y)$, after a forward in time integration for $t=289 \, t_{\mathrm{u}}$ of the orbits considered in Fig.~\ref{fig:escapes}(c). Stars colored in purple in (a) and (c) correspond to orbits undergoing $N_{\mathrm{e}} \leq 2$ escapes from the central region for times up to  $t = 1071 \, t_{\mathrm{u}}$, while stars colored in cyan in (b) and (c) correspond to orbits with $N_{\mathrm{e}} > 2$. In (c) the sets of points from (a) and (b) are plotted together. In all panels the energetically forbidden regions are colored in gray. }
	\label{fig:morphology_1}
\end{figure}

Following the same color convention used in  Fig.~\ref{fig:escapes}(c), we  plot in Fig.~\ref{fig:morphology_1} the positions, after $t=289 \, t_{\mathrm{u}}$ ($\approx 1.35 \mbox{Gyrs}$), of stars whose orbits have as initial conditions the ones considered  in Fig.~\ref{fig:escapes}. We note that for obtaining the results of Fig.~\ref{fig:morphology_1} we did not integrate the orbits up to $t=1071 \, t_{\mathrm{u}}$ ($\approx 5 \mbox{Gyrs}$) as was done in Fig.~\ref{fig:escapes}, because this long integration results to a large portion of the orbits leaving the central region of the system and traveling very far away in the exterior region, not supporting in this way the formation of clear morphological structures. On the other hand, as we see in Fig.~\ref{fig:morphology_1}, for $t=289 \, t_{\mathrm{u}}$  the majority of the considered  orbits remain in the proximity of the systems origin, supporting some well defined morphological features. In Fig.~\ref{fig:morphology_1}(a) we plot in purple the positions (at $t=289 \, t_{\mathrm{u}}$) of the stars whose orbits undergo $N_{\mathrm{e}} \leq 2$ escapes from the central region until $t = 1071 \, t_{\mathrm{u}}$, while orbits with $N_{\mathrm{e}} > 2$ are shown in cyan in Fig.~\ref{fig:morphology_1}(b). In Fig.~\ref{fig:morphology_1}(c)  we superimpose the populations of stars presented in Figs.~\ref{fig:morphology_1}(a)  and (b). In all panels of Fig.~\ref{fig:morphology_1} the energetically forbidden regions are depicted in gray. We note that due to the symmetry of the stable and unstable manifolds  with respect to the $p_y=0$ (Fig.~\ref{fig:LDs_manifolds}), we present in Fig.~\ref{fig:morphology_1} results  only for the forward time evolution of orbits, which means that these orbits are influenced by the morphology of the stable manifolds. 

A main difference between Figs.~\ref{fig:morphology_1}(a)  and (b) is that the number of stars experiencing many escapes from the central region (and consequently several reentries to that region) [Fig.~\ref{fig:morphology_1}(b)] is  much smaller then the ones plotted in Fig.~\ref{fig:morphology_1}(a). This difference means that the majority of studied orbits escape only a few times from the central region.  Furthermore, these orbits'  final positions,   depicted in Fig.~\ref{fig:morphology_1}(a), can be to considerably larger distances from the center with respect to the stars in Fig.~\ref{fig:morphology_1}(b). On the other hand, orbits which enter the central region and escape from it many times [Fig.~\ref{fig:morphology_1}(b)] remain in the vicinity of the galaxy's central part supporting the creation of morphological structures in that region.

\begin{figure}[tbhp!]
	\centering
		\text{\footnotesize{~~~~~~~~~~~~~~~1--1~~~~~~~~~~~~~~~~2--2~~~~~~~~~~~~homoclinic~~~~~~~~~~~1--2~~~~~~~~~~~~~~~~2--1~~~~~~~~~~~heteroclinic}}\par\smallskip
		\includegraphics[width=\textwidth]{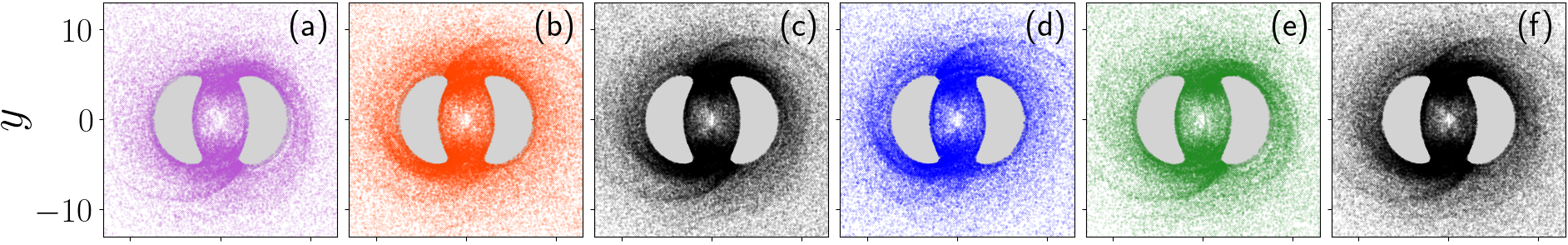}    
		\includegraphics[width=\textwidth]{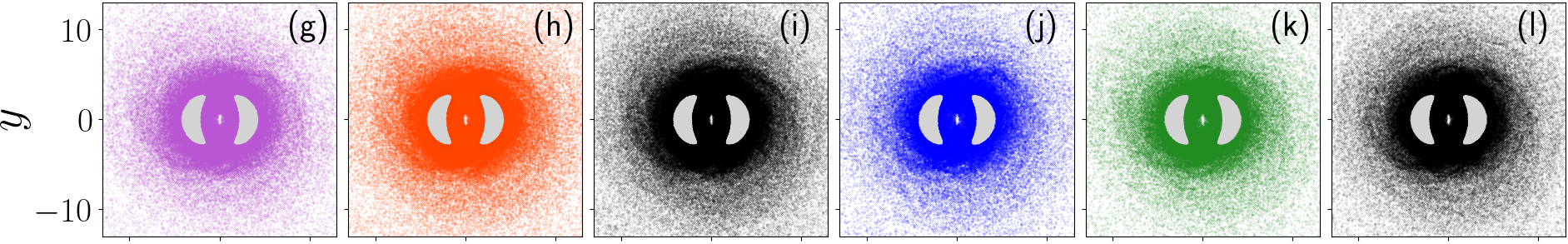}    
		\includegraphics[width=\textwidth]{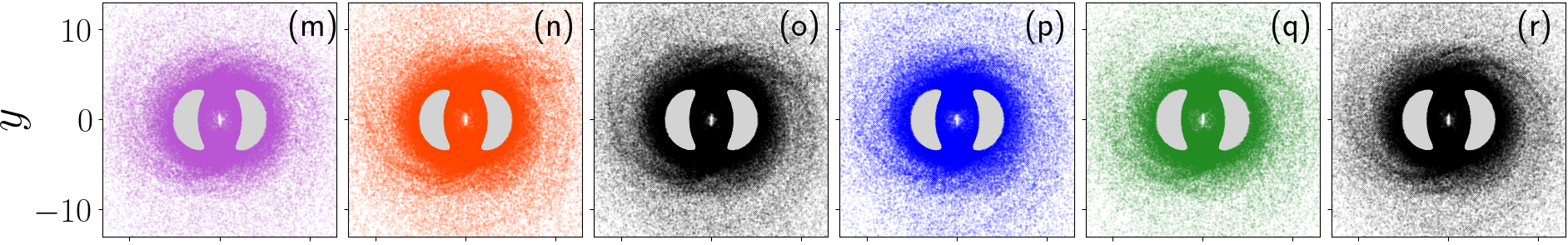}    
		\includegraphics[width=\textwidth]{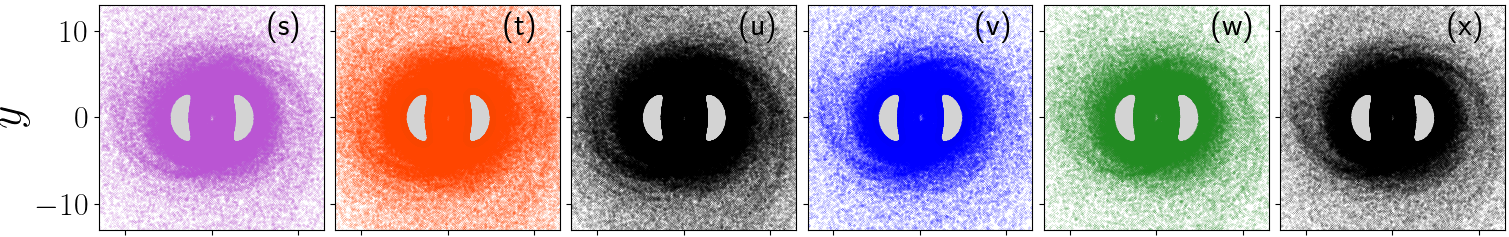}    
		\includegraphics[width=\textwidth]{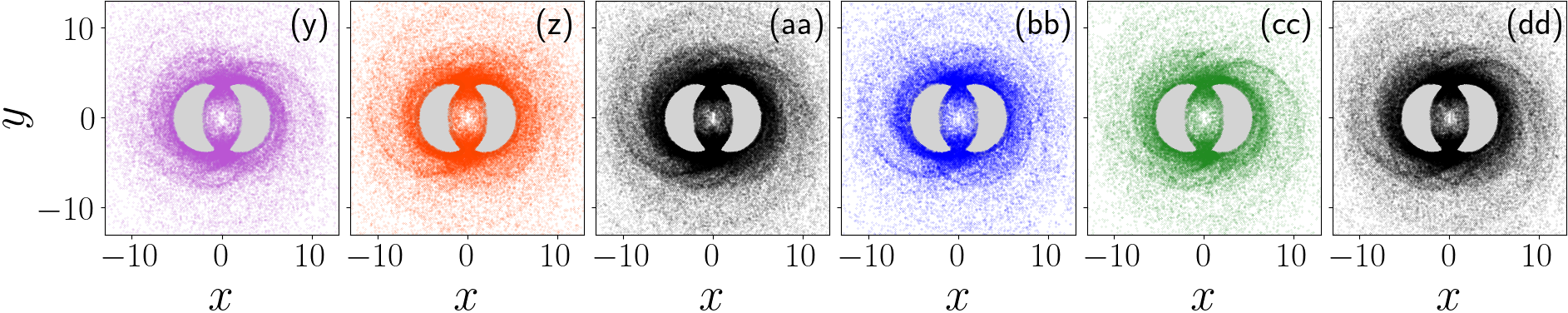}    
	\caption{Final positions of stars in system's \eqref{eq:2DHam} configuration space, after a forward in time integration for $t=289 \, t_{\mathrm{u}}$ of  orbits whose initial conditions correspond to different regions of the OFMs of Fig.~\ref{fig:5OFM}. Plots correspond to the following parameter arrangements: 
		[(a)--(f)] $E=-3.91$ and $\Omega_{\mathrm{b}}=45\,\mbox{km} \,\mbox{s}^{-1}\, \mbox{kpc}^{-1}$,
		[(g)--(l)] $E=-4.37$ and $\Omega_{\mathrm{b}}=70\,\mbox{km} \,\mbox{s}^{-1}\, \mbox{kpc}^{-1}$,
		[(m)--(r)] $E=-4.2$ and $\Omega_{\mathrm{b}}=60\,\mbox{km} \,\mbox{s}^{-1}\, \mbox{kpc}^{-1}$,
		[(s)--(x)] $E=-4.15$ and $\Omega_{\mathrm{b}}=60\,\mbox{km} \,\mbox{s}^{-1}\, \mbox{kpc}^{-1}$, and 
		[(y)--(dd)] $E=-4.235$  and $\Omega_{\mathrm{b}}=60\,\mbox{km} \,\mbox{s}^{-1}\, \mbox{kpc}^{-1}$.
		Results presented in different columns refer to different combinations of the OFM indices: (first column, purple points) 1-1, (second column, red points) 2-2, (fourth column, blue points) 1-2 and (fifth column, green points) 2-1. Results  from homoclinic (heteroclinic) orbits are obtained by the superposition of the points in the panels of the first and second (forth and fifth) columns and are presented in black in the third (sixth) column. The particular orbit classification presented in each column is also mentioned at the top of the panels of the upper row. In all panels the energetically forbidden regions are colored in gray.}
	\label{fig:morphology_2}
\end{figure}

In order to reveal   morphological features associated with the presence of homoclinic and heteroclinic orbits (Fig.~\ref{fig:5homhet}), as well as with orbits belonging to different regions in the OFMs (Fig.~\ref{fig:5OFM}), we implement a similar approach to the one used for the creation of Fig.~\ref{fig:morphology_1} for the cases  presented in Figs.~\ref{fig:5LDplots}, \ref{fig:5homhet} and \ref{fig:5OFM}, which correspond to various $E$ and $\Omega_{\mathrm{b}}$ values, as well as different sizes of the openings between the energetically  forbidden regions. The outcomes of this process are presented in Fig.~\ref{fig:morphology_2}. In particular,  we present in Fig.~\ref{fig:morphology_2} results obtained for the following parameter arrangements of Hamiltonian \eqref{eq:2DHam}: 
$E=-3.91$ and $\Omega_{\mathrm{b}}=45\,\mbox{km} \,\mbox{s}^{-1}\, \mbox{kpc}^{-1}$ [Figs.~\ref{fig:morphology_2}(a)--(f)],
$E=-4.37$ and $\Omega_{\mathrm{b}}=70\,\mbox{km} \,\mbox{s}^{-1}\, \mbox{kpc}^{-1}$ [Figs.~\ref{fig:morphology_2}(g)--(l)],
$E=-4.2$ and $\Omega_{\mathrm{b}}=60\,\mbox{km} \,\mbox{s}^{-1}\, \mbox{kpc}^{-1}$ [Figs.~\ref{fig:morphology_2}(m)--(r)],
$E=-4.15$ and $\Omega_{\mathrm{b}}=60\,\mbox{km} \,\mbox{s}^{-1}\, \mbox{kpc}^{-1}$ [Figs.~\ref{fig:morphology_2}(s)--(x)], and 
$E=-4.235$  and $\Omega_{\mathrm{b}}=60\,\mbox{km} \,\mbox{s}^{-1}\, \mbox{kpc}^{-1}$ [Figs.~\ref{fig:morphology_2}(y)--(dd)].
We note that the results  presented in the upper three rows of Fig.~\ref{fig:morphology_2} correspond to cases practically having the same openings size. These cases were respectively presented in panels (a), (c) and (b) of Figs.~\ref{fig:5LDplots}, \ref{fig:5homhet} and \ref{fig:5OFM}. The cases shown in the bottom three rows of Fig.~\ref{fig:morphology_2} have the same pattern speed $\Omega_{\mathrm{b}}=60\,\mbox{km} \,\mbox{s}^{-1}\, \mbox{kpc}^{-1}$, and  were respectively presented in panels (b), (d) and (e) of Figs.~\ref{fig:5LDplots}, \ref{fig:5homhet} and \ref{fig:5OFM}.

From the examination of the results presented in Fig.~\ref{fig:morphology_2} we see that the morphologies created by orbits corresponding to OFM  combinations 1-1 and 2-1 are  very similar to each other. The same is true for the morphologies obtained from the 2-2 and 1-2 orbits. These similarities denote that  the opening from which orbits escape from the central region in the future is mainly determining the morphological features that these orbits  support. In addition, since features created by the 1-1 and 2-2  orbits (which together correspond to  homoclinic orbits)  closely resemble  the features formed by the 2-1 and 1-2 orbits (whose  combination correspond to heteroclinic orbits), we understand that homoclinic  (plots in the third column of  Fig.~\ref{fig:morphology_2}), and heteroclinic orbits (sixth column of  Fig.~\ref{fig:morphology_2}) support similar morphological features.

With respect to the effect of the pattern speed on the different morphological structures, we note that the lowest considered pattern speed of $\Omega_{\mathrm{b}}=45\,\mbox{km} \,\mbox{s}^{-1}\, \mbox{kpc}^{-1}$ [Figs.~\ref{fig:morphology_2}(a)--(f)] results to the creation of spiral-like features, which are more pronounced close to the system's openings. On the other hand, in the case with the largest pattern speed $\Omega_{\mathrm{b}}=70\,\mbox{km} \,\mbox{s}^{-1}\, \mbox{kpc}^{-1}$ [Figs.~\ref{fig:morphology_2}(g)--(l)]  we observe more circular structures surrounding the energetically forbidden regions. The spiral-like features resulting from $\Omega_{\mathrm{b}}=60\,\mbox{km} \,\mbox{s}^{-1}\, \mbox{kpc}^{-1}$  for an energy value which allows the same openings' size  as in the previous two cases [Figs.~\ref{fig:morphology_2}(m)--(r)] are not as clear as the ones in Figs.~\ref{fig:morphology_2}(a)--(f),  but also not as circular as in Figs.~\ref{fig:morphology_2}(g)--(l).

We have already observed from the comparison of Figs.~\ref{fig:5OFM}(a) and (c), that increasing the pattern speed for cases where the size of the openings remain practically the same, diminishes the sensitivity of  the orbits dynamical evolution on the location of their initial conditions. In Fig.~\ref{fig:morphology_2} we also see that an increase in the mixing of the various dynamical behaviors in the OFM leads to the creation of more pronounced spiral-like morphologies  [Figs.~\ref{fig:morphology_2}(a)--(f)]. 

Comparing the results of the cases presented in bottom three rows of Fig.~\ref{fig:morphology_2},  all of which have the same pattern speed $\Omega_{\mathrm{b}}=60\,\mbox{km} \,\mbox{s}^{-1}\, \mbox{kpc}^{-1}$, we see that in the case corresponding to the largest energy $E=-4.15$ [Figs.~\ref{fig:morphology_2}(s)--(x)] more stars are present near  the central region and around the energetically forbidden areas of the galaxy, since in this case the larger openings allow more orbits to move in and out of the central region. It is  worth noting that in the case with the  lowest energy $E=-4.235$ [Figs.~\ref{fig:morphology_2}(y)--(dd)], where the size of the openings is quite small, some well defined spiral-like structures are observed around the energetically forbidden regions.

\section{Summary}
\label{sec:summary}

We investigated the influence of stable and unstable manifolds on the orbital behavior of a 2D Hamiltonian system \eqref{eq:2DHam} that describes  the motion of a star in the equatorial plane of a simple rotating galactic potential. In particular, we were interested in the way that orbits starting in  the central part of the model escape to the exterior regions. For particular ranges of energies these escapes can happen through openings between energetically forbidden regions located in the area of the Lagrange points $L_1$ and $L_2$. Our investigation was based on the extensive numerical propagation of ensembles of orbits, both forward and backward in time, the computation of their LDs, along with the creation of several OFMs for various values of the model's parameters (pattern speed $\Omega_{\mathrm{b}}$ and energy $E$). From the computation of the LDs we were able to reveal the structure and location of the system's stable and unstable manifolds (Fig.~\ref{fig:LDs_manifolds}), while OFMs allowed us to identify sets of orbits having similar dynamical traits by keeping track of the specific openings from which orbits escaped in the future from (or entered in the past in) the system's central region.

We showed that various characteristics of the future evolution of orbits, like the time $t_{\mathrm{e}}$ of the first escape from the central region, the total time $t_{\mathrm{d}}$ that orbits spent at the exterior parts of the system, as well as the number $N_{\mathrm{e}}$ of successive exits from the central region, are influenced by the morphology of this stable manifolds (Fig.~\ref{fig:escapes}). 

We also examined the influence of the manifolds on the past and future evolution of orbits, by considering the properties of ensemble of initial conditions leading to similar orbital traits. In particular, we considered homoclinic (heteroclinic) orbits, i.e.~orbits which escape from the system's central region through the same (different) opening when they are integrated both forward and backward in time (Fig.~\ref{fig:5homhet}), as well as orbits corresponding to different arrangements of the OFM indices (Figs.~\ref{fig:5OFM} and \ref{fig:OFMzoom}) by taking into account all   possible combinations of passes through the different openings. We investigated how the mixing of phase space regions of initial conditions leading to homoclinic and heteroclinic, as well as orbits with different OFM characterizations, change when the model's pattern speed and energy were altered so that the openings' size remained practically constant, or when $E$ changed for a fixed value of $\Omega_{\mathrm{b}}$. A main outcome of our analysis was that the sensitivity of the observed orbital behaviors on the location of the orbit's initial condition was decreased for cases having similar opening sizes when $\Omega_{\mathrm{b}}$ grew, or when the openings between the energetically forbidden regions decreased in size for fixed $\Omega_{\mathrm{b}}$ values.

Analyzing in Figs.~\ref{fig:morphology_1} and \ref{fig:morphology_2} the morphological structures supported by different types of orbits in the system's configuration space, we observed no significant differences between the features associated with homoclinic and heteroclinic orbits, when these orbits were integrated forward in time. This behavior is a consequence of the fact that the future evolution  of escaping orbits is mainly defined by the opening from which they escape and not by the opening from which they entered in the central region in the past, as well as, by the observation that orbits with 1-1 and 2-1 indices in the OFM create similar patterns, something which is also true for the 2-2 and 1-2 orbits. Furthermore, we saw that an increase in the system's pattern speed results in  orbits supporting more circular-like structures, while lower $\Omega_{\mathrm{b}}$ values favor more spiral-like morphologies.

Our study constitutes, to the best of our knowledge, the first attempt to analyze the orbital behavior of galactic models based on the computation of LDs and on the creation of the OFM. We believe that these are very powerful tools which give us the ability to understand in more detail the influence of different aspects of the system on its orbital behavior, and eventually on the galactic morphologies the such behaviors support. Thus, we hope that our work will initiate further applications of LDs and OFMs in galactic dynamics studies.

\section*{Acknowledgements}

We thank the Center for High Performance Computing (CHPC) of South Africa and the High Performance Computing facility of the University of Cape Town (UCT) for providing computational resources for this work, as well as Dr.~Matthaios Katsanikas for valuable discussions. D.~T.~thanks UCT and the Harry Crossley Foundation for their financial support.





\bibliography{Theron_Skokos_Galactic_model_bib}

\begin{thebibliography}{56}%
\makeatletter
\providecommand \@ifxundefined [1]{%
 \@ifx{#1\undefined}
}%
\providecommand \@ifnum [1]{%
 \ifnum #1\expandafter \@firstoftwo
 \else \expandafter \@secondoftwo
 \fi
}%
\providecommand \@ifx [1]{%
 \ifx #1\expandafter \@firstoftwo
 \else \expandafter \@secondoftwo
 \fi
}%
\providecommand \natexlab [1]{#1}%
\providecommand \enquote  [1]{``#1''}%
\providecommand \bibnamefont  [1]{#1}%
\providecommand \bibfnamefont [1]{#1}%
\providecommand \citenamefont [1]{#1}%
\providecommand \href@noop [0]{\@secondoftwo}%
\providecommand \href [0]{\begingroup \@sanitize@url \@href}%
\providecommand \@href[1]{\@@startlink{#1}\@@href}%
\providecommand \@@href[1]{\endgroup#1\@@endlink}%
\providecommand \@sanitize@url [0]{\catcode `\\12\catcode `\$12\catcode
  `\&12\catcode `\#12\catcode `\^12\catcode `\_12\catcode `\%12\relax}%
\providecommand \@@startlink[1]{}%
\providecommand \@@endlink[0]{}%
\providecommand \url  [0]{\begingroup\@sanitize@url \@url }%
\providecommand \@url [1]{\endgroup\@href {#1}{\urlprefix }}%
\providecommand \urlprefix  [0]{URL }%
\providecommand \Eprint [0]{\href }%
\providecommand \doibase [0]{https://doi.org/}%
\providecommand \selectlanguage [0]{\@gobble}%
\providecommand \bibinfo  [0]{\@secondoftwo}%
\providecommand \bibfield  [0]{\@secondoftwo}%
\providecommand \translation [1]{[#1]}%
\providecommand \BibitemOpen [0]{}%
\providecommand \bibitemStop [0]{}%
\providecommand \bibitemNoStop [0]{.\EOS\space}%
\providecommand \EOS [0]{\spacefactor3000\relax}%
\providecommand \BibitemShut  [1]{\csname bibitem#1\endcsname}%
\let\auto@bib@innerbib\@empty
\bibitem [{\citenamefont {Marinova}\ and\ \citenamefont {Jogee}(2007)}]{MJ07}%
  \BibitemOpen
  \bibfield  {author} {\bibinfo {author} {\bibfnamefont {I.}~\bibnamefont
  {Marinova}}\ and\ \bibinfo {author} {\bibfnamefont {S.}~\bibnamefont
  {Jogee}},\ }\href@noop {} {\bibfield  {journal} {\bibinfo  {journal} {The
  Astrophysical Journal}\ }\textbf {\bibinfo {volume} {\textbf{659}}},\
  \bibinfo {pages} {1176} (\bibinfo {year} {2007})}\BibitemShut {NoStop}%
\bibitem [{\citenamefont {Barazza}\ \emph {et~al.}(2008)\citenamefont
  {Barazza}, \citenamefont {Jogee},\ and\ \citenamefont {Marinova}}]{BJM08}%
  \BibitemOpen
  \bibfield  {author} {\bibinfo {author} {\bibfnamefont {F.~D.}\ \bibnamefont
  {Barazza}}, \bibinfo {author} {\bibfnamefont {S.}~\bibnamefont {Jogee}},\
  and\ \bibinfo {author} {\bibfnamefont {I.}~\bibnamefont {Marinova}},\
  }\href@noop {} {\bibfield  {journal} {\bibinfo  {journal} {The Astrophysical
  Journal}\ }\textbf {\bibinfo {volume} {\textbf{675}}},\ \bibinfo {pages}
  {1194} (\bibinfo {year} {2008})}\BibitemShut {NoStop}%
\bibitem [{\citenamefont {Elmegreen}\ and\ \citenamefont
  {Elmegreen}(1982)}]{EE82}%
  \BibitemOpen
  \bibfield  {author} {\bibinfo {author} {\bibfnamefont {D.~M.}\ \bibnamefont
  {Elmegreen}}\ and\ \bibinfo {author} {\bibfnamefont {B.~G.}\ \bibnamefont
  {Elmegreen}},\ }\href@noop {} {\bibfield  {journal} {\bibinfo  {journal}
  {Monthly Notices of the Royal Astronomical Society}\ }\textbf {\bibinfo
  {volume} {\textbf{201}}},\ \bibinfo {pages} {1021} (\bibinfo {year}
  {1982})}\BibitemShut {NoStop}%
\bibitem [{\citenamefont {Eskridge}\ \emph {et~al.}(2000)\citenamefont
  {Eskridge}, \citenamefont {Frogel}, \citenamefont {Pogge}, \citenamefont
  {Quillen}, \citenamefont {Davies}, \citenamefont {DePoy}, \citenamefont
  {Houdashelt}, \citenamefont {Kuchinski}, \citenamefont {Ramírez},
  \citenamefont {Sellgren}, \citenamefont {Terndrup},\ and\ \citenamefont
  {Tiede}}]{EFPQDDHKRSTT00}%
  \BibitemOpen
  \bibfield  {author} {\bibinfo {author} {\bibfnamefont {P.~B.}\ \bibnamefont
  {Eskridge}}, \bibinfo {author} {\bibfnamefont {J.~A.}\ \bibnamefont
  {Frogel}}, \bibinfo {author} {\bibfnamefont {R.~W.}\ \bibnamefont {Pogge}},
  \bibinfo {author} {\bibfnamefont {A.~C.}\ \bibnamefont {Quillen}}, \bibinfo
  {author} {\bibfnamefont {R.~L.}\ \bibnamefont {Davies}}, \bibinfo {author}
  {\bibfnamefont {D.~L.}\ \bibnamefont {DePoy}}, \bibinfo {author}
  {\bibfnamefont {M.~L.}\ \bibnamefont {Houdashelt}}, \bibinfo {author}
  {\bibfnamefont {L.~E.}\ \bibnamefont {Kuchinski}}, \bibinfo {author}
  {\bibfnamefont {S.~V.}\ \bibnamefont {Ramírez}}, \bibinfo {author}
  {\bibfnamefont {K.}~\bibnamefont {Sellgren}}, \bibinfo {author}
  {\bibfnamefont {D.~M.}\ \bibnamefont {Terndrup}},\ and\ \bibinfo {author}
  {\bibfnamefont {G.~P.}\ \bibnamefont {Tiede}},\ }\href@noop {} {\bibfield
  {journal} {\bibinfo  {journal} {The Astronomical Journal}\ }\textbf {\bibinfo
  {volume} {\textbf{119}}},\ \bibinfo {pages} {536} (\bibinfo {year}
  {2000})}\BibitemShut {NoStop}%
\bibitem [{\citenamefont {Buta}(1986)}]{B86}%
  \BibitemOpen
  \bibfield  {author} {\bibinfo {author} {\bibfnamefont {R.}~\bibnamefont
  {Buta}},\ }\href@noop {} {\bibfield  {journal} {\bibinfo  {journal} {The
  Astrophysical Journal Supplement Series}\ }\textbf {\bibinfo {volume}
  {\textbf{61}}},\ \bibinfo {pages} {609} (\bibinfo {year} {1986})}\BibitemShut
  {NoStop}%
\bibitem [{\citenamefont {Buta}(1995)}]{B95}%
  \BibitemOpen
  \bibfield  {author} {\bibinfo {author} {\bibfnamefont {R.}~\bibnamefont
  {Buta}},\ }\href@noop {} {\bibfield  {journal} {\bibinfo  {journal} {The
  Astrophysical Journal Supplement Series}\ }\textbf {\bibinfo {volume}
  {\textbf{96}}},\ \bibinfo {pages} {39} (\bibinfo {year} {1995})}\BibitemShut
  {NoStop}%
\bibitem [{\citenamefont {Athanassoula}(1984)}]{A84}%
  \BibitemOpen
  \bibfield  {author} {\bibinfo {author} {\bibfnamefont {E.}~\bibnamefont
  {Athanassoula}},\ }\href@noop {} {\bibfield  {journal} {\bibinfo  {journal}
  {Physics Reports}\ }\textbf {\bibinfo {volume} {\textbf{114}}},\ \bibinfo
  {pages} {319} (\bibinfo {year} {1984})}\BibitemShut {NoStop}%
\bibitem [{\citenamefont {Sellwood}\ and\ \citenamefont
  {Wilkinson}(1993)}]{SW93}%
  \BibitemOpen
  \bibfield  {author} {\bibinfo {author} {\bibfnamefont {J.~A.}\ \bibnamefont
  {Sellwood}}\ and\ \bibinfo {author} {\bibfnamefont {A.}~\bibnamefont
  {Wilkinson}},\ }\href@noop {} {\bibfield  {journal} {\bibinfo  {journal}
  {Reports on Progress in Physics}\ }\textbf {\bibinfo {volume}
  {\textbf{56}}},\ \bibinfo {pages} {173} (\bibinfo {year} {1993})}\BibitemShut
  {NoStop}%
\bibitem [{\citenamefont {Athanassoula}\ \emph {et~al.}(1983)\citenamefont
  {Athanassoula}, \citenamefont {Bienaym{\'e}}, \citenamefont {Martinet},\ and\
  \citenamefont {Pfenniger}}]{ABMP83}%
  \BibitemOpen
  \bibfield  {author} {\bibinfo {author} {\bibfnamefont {E.}~\bibnamefont
  {Athanassoula}}, \bibinfo {author} {\bibfnamefont {O.}~\bibnamefont
  {Bienaym{\'e}}}, \bibinfo {author} {\bibfnamefont {L.}~\bibnamefont
  {Martinet}},\ and\ \bibinfo {author} {\bibfnamefont {D.}~\bibnamefont
  {Pfenniger}},\ }\href@noop {} {\bibfield  {journal} {\bibinfo  {journal}
  {Astronomy \& Astrophysics}\ }\textbf {\bibinfo {volume} {\textbf{127}}},\
  \bibinfo {pages} {349} (\bibinfo {year} {1983})}\BibitemShut {NoStop}%
\bibitem [{\citenamefont {Contopoulos}\ and\ \citenamefont
  {Grosbøl}(1989)}]{CG89}%
  \BibitemOpen
  \bibfield  {author} {\bibinfo {author} {\bibfnamefont {G.}~\bibnamefont
  {Contopoulos}}\ and\ \bibinfo {author} {\bibfnamefont {P.}~\bibnamefont
  {Grosbøl}},\ }\href@noop {} {\bibfield  {journal} {\bibinfo  {journal} {The
  Astronomy and Astrophysics Review}\ }\textbf {\bibinfo {volume}
  {\textbf{1}}},\ \bibinfo {pages} {261} (\bibinfo {year} {1989})}\BibitemShut
  {NoStop}%
\bibitem [{\citenamefont {Skokos}\ \emph
  {et~al.}(2002{\natexlab{a}})\citenamefont {Skokos}, \citenamefont {Patsis},\
  and\ \citenamefont {Athanassoula}}]{SPA02a}%
  \BibitemOpen
  \bibfield  {author} {\bibinfo {author} {\bibfnamefont {C.}~\bibnamefont
  {Skokos}}, \bibinfo {author} {\bibfnamefont {P.~A.}\ \bibnamefont {Patsis}},\
  and\ \bibinfo {author} {\bibfnamefont {E.}~\bibnamefont {Athanassoula}},\
  }\href@noop {} {\bibfield  {journal} {\bibinfo  {journal} {Monthly Notices of
  the Royal Astronomical Society}\ }\textbf {\bibinfo {volume}
  {\textbf{333}}},\ \bibinfo {pages} {847} (\bibinfo {year}
  {2002}{\natexlab{a}})}\BibitemShut {NoStop}%
\bibitem [{\citenamefont {Skokos}\ \emph
  {et~al.}(2002{\natexlab{b}})\citenamefont {Skokos}, \citenamefont {Patsis},\
  and\ \citenamefont {Athanassoula}}]{SPA02b}%
  \BibitemOpen
  \bibfield  {author} {\bibinfo {author} {\bibfnamefont {C.}~\bibnamefont
  {Skokos}}, \bibinfo {author} {\bibfnamefont {P.~A.}\ \bibnamefont {Patsis}},\
  and\ \bibinfo {author} {\bibfnamefont {E.}~\bibnamefont {Athanassoula}},\
  }\href@noop {} {\bibfield  {journal} {\bibinfo  {journal} {Monthly Notices of
  the Royal Astronomical Society}\ }\textbf {\bibinfo {volume}
  {\textbf{333}}},\ \bibinfo {pages} {861} (\bibinfo {year}
  {2002}{\natexlab{b}})}\BibitemShut {NoStop}%
\bibitem [{\citenamefont {Patsis}\ \emph {et~al.}(2002)\citenamefont {Patsis},
  \citenamefont {Skokos},\ and\ \citenamefont {Athanassoula}}]{PSA02}%
  \BibitemOpen
  \bibfield  {author} {\bibinfo {author} {\bibfnamefont {P.~A.}\ \bibnamefont
  {Patsis}}, \bibinfo {author} {\bibfnamefont {C.}~\bibnamefont {Skokos}},\
  and\ \bibinfo {author} {\bibfnamefont {E.}~\bibnamefont {Athanassoula}},\
  }\href@noop {} {\bibfield  {journal} {\bibinfo  {journal} {Monthly Notices of
  the Royal Astronomical Society}\ }\textbf {\bibinfo {volume}
  {\textbf{337}}},\ \bibinfo {pages} {578} (\bibinfo {year}
  {2002})}\BibitemShut {NoStop}%
\bibitem [{\citenamefont {Patsis}\ \emph {et~al.}(2003)\citenamefont {Patsis},
  \citenamefont {Skokos},\ and\ \citenamefont {Athanassoula}}]{PSA03}%
  \BibitemOpen
  \bibfield  {author} {\bibinfo {author} {\bibfnamefont {P.~A.}\ \bibnamefont
  {Patsis}}, \bibinfo {author} {\bibfnamefont {C.}~\bibnamefont {Skokos}},\
  and\ \bibinfo {author} {\bibfnamefont {E.}~\bibnamefont {Athanassoula}},\
  }\href@noop {} {\bibfield  {journal} {\bibinfo  {journal} {Monthly Notices of
  the Royal Astronomical Society}\ }\textbf {\bibinfo {volume}
  {\textbf{342}}},\ \bibinfo {pages} {69} (\bibinfo {year} {2003})}\BibitemShut
  {NoStop}%
\bibitem [{\citenamefont {Kaufmann}\ and\ \citenamefont
  {Contopoulos}(1996)}]{KC96}%
  \BibitemOpen
  \bibfield  {author} {\bibinfo {author} {\bibfnamefont {D.~E.}\ \bibnamefont
  {Kaufmann}}\ and\ \bibinfo {author} {\bibfnamefont {G.}~\bibnamefont
  {Contopoulos}},\ }\href@noop {} {\bibfield  {journal} {\bibinfo  {journal}
  {Astronomy \& Astrophysics}\ }\textbf {\bibinfo {volume} {\textbf{309}}},\
  \bibinfo {pages} {381} (\bibinfo {year} {1996})}\BibitemShut {NoStop}%
\bibitem [{\citenamefont {Wozniak}\ and\ \citenamefont
  {Pfenniger}(1999)}]{WP99}%
  \BibitemOpen
  \bibfield  {author} {\bibinfo {author} {\bibfnamefont {H.}~\bibnamefont
  {Wozniak}}\ and\ \bibinfo {author} {\bibfnamefont {D.}~\bibnamefont
  {Pfenniger}},\ }\href@noop {} {\bibfield  {journal} {\bibinfo  {journal}
  {Celestial Mechanics and Dynamical Astronomy}\ }\textbf {\bibinfo {volume}
  {\textbf{73}}},\ \bibinfo {pages} {149} (\bibinfo {year} {1999})}\BibitemShut
  {NoStop}%
\bibitem [{\citenamefont {Contopoulos}\ and\ \citenamefont
  {Harsoula}(2013)}]{CH13}%
  \BibitemOpen
  \bibfield  {author} {\bibinfo {author} {\bibfnamefont {G.}~\bibnamefont
  {Contopoulos}}\ and\ \bibinfo {author} {\bibfnamefont {M.}~\bibnamefont
  {Harsoula}},\ }\href@noop {} {\bibfield  {journal} {\bibinfo  {journal}
  {Monthly Notices of the Royal Astronomical Society}\ }\textbf {\bibinfo
  {volume} {\textbf{436}}},\ \bibinfo {pages} {1201} (\bibinfo {year}
  {2013})}\BibitemShut {NoStop}%
\bibitem [{\citenamefont {Tsigaridi}\ and\ \citenamefont
  {Patsis}(2015)}]{TP15}%
  \BibitemOpen
  \bibfield  {author} {\bibinfo {author} {\bibfnamefont {L.}~\bibnamefont
  {Tsigaridi}}\ and\ \bibinfo {author} {\bibfnamefont {P.~A.}\ \bibnamefont
  {Patsis}},\ }\href@noop {} {\bibfield  {journal} {\bibinfo  {journal}
  {Monthly Notices of the Royal Astronomical Society}\ }\textbf {\bibinfo
  {volume} {\textbf{448}}},\ \bibinfo {pages} {3081} (\bibinfo {year}
  {2015})}\BibitemShut {NoStop}%
\bibitem [{\citenamefont {Romero-G{\'o}mez}\ \emph {et~al.}(2006)\citenamefont
  {Romero-G{\'o}mez}, \citenamefont {Masdemont}, \citenamefont {Athanassoula},\
  and\ \citenamefont {Garc{\'\i}a-G{\'o}mez}}]{RMAG06}%
  \BibitemOpen
  \bibfield  {author} {\bibinfo {author} {\bibfnamefont {M.}~\bibnamefont
  {Romero-G{\'o}mez}}, \bibinfo {author} {\bibfnamefont {J.}~\bibnamefont
  {Masdemont}}, \bibinfo {author} {\bibfnamefont {E.}~\bibnamefont
  {Athanassoula}},\ and\ \bibinfo {author} {\bibfnamefont {C.}~\bibnamefont
  {Garc{\'\i}a-G{\'o}mez}},\ }\href@noop {} {\bibfield  {journal} {\bibinfo
  {journal} {Astronomy \& Astrophysics}\ }\textbf {\bibinfo {volume}
  {\textbf{453}}},\ \bibinfo {pages} {39} (\bibinfo {year} {2006})}\BibitemShut
  {NoStop}%
\bibitem [{\citenamefont {Romero-G{\'o}mez}\ \emph {et~al.}(2007)\citenamefont
  {Romero-G{\'o}mez}, \citenamefont {Athanassoula}, \citenamefont {Masdemont},\
  and\ \citenamefont {Garc{\'\i}a-G{\'o}mez}}]{RAMG07}%
  \BibitemOpen
  \bibfield  {author} {\bibinfo {author} {\bibfnamefont {M.}~\bibnamefont
  {Romero-G{\'o}mez}}, \bibinfo {author} {\bibfnamefont {E.}~\bibnamefont
  {Athanassoula}}, \bibinfo {author} {\bibfnamefont {J.}~\bibnamefont
  {Masdemont}},\ and\ \bibinfo {author} {\bibfnamefont {C.}~\bibnamefont
  {Garc{\'\i}a-G{\'o}mez}},\ }\href@noop {} {\bibfield  {journal} {\bibinfo
  {journal} {Astronomy \& Astrophysics}\ }\textbf {\bibinfo {volume}
  {\textbf{472}}},\ \bibinfo {pages} {63} (\bibinfo {year} {2007})}\BibitemShut
  {NoStop}%
\bibitem [{\citenamefont {Athanassoula}\ \emph
  {et~al.}(2009{\natexlab{a}})\citenamefont {Athanassoula}, \citenamefont
  {Romero-G{\'o}mez},\ and\ \citenamefont {Masdemont}}]{ARM09}%
  \BibitemOpen
  \bibfield  {author} {\bibinfo {author} {\bibfnamefont {E.}~\bibnamefont
  {Athanassoula}}, \bibinfo {author} {\bibfnamefont {M.}~\bibnamefont
  {Romero-G{\'o}mez}},\ and\ \bibinfo {author} {\bibfnamefont {J.}~\bibnamefont
  {Masdemont}},\ }\href@noop {} {\bibfield  {journal} {\bibinfo  {journal}
  {Monthly Notices of the Royal Astronomical Society}\ }\textbf {\bibinfo
  {volume} {\textbf{394}}},\ \bibinfo {pages} {67} (\bibinfo {year}
  {2009}{\natexlab{a}})}\BibitemShut {NoStop}%
\bibitem [{\citenamefont {Athanassoula}\ \emph
  {et~al.}(2009{\natexlab{b}})\citenamefont {Athanassoula}, \citenamefont
  {Romero-G{\'o}mez}, \citenamefont {Bosma},\ and\ \citenamefont
  {Masdemont}}]{ARBM09}%
  \BibitemOpen
  \bibfield  {author} {\bibinfo {author} {\bibfnamefont {E.}~\bibnamefont
  {Athanassoula}}, \bibinfo {author} {\bibfnamefont {M.}~\bibnamefont
  {Romero-G{\'o}mez}}, \bibinfo {author} {\bibfnamefont {A.}~\bibnamefont
  {Bosma}},\ and\ \bibinfo {author} {\bibfnamefont {J.}~\bibnamefont
  {Masdemont}},\ }\href@noop {} {\bibfield  {journal} {\bibinfo  {journal}
  {Monthly Notices of the Royal Astronomical Society}\ }\textbf {\bibinfo
  {volume} {\textbf{400}}},\ \bibinfo {pages} {1706} (\bibinfo {year}
  {2009}{\natexlab{b}})}\BibitemShut {NoStop}%
\bibitem [{\citenamefont {Madrid}\ and\ \citenamefont {Mancho}(2009)}]{MM09}%
  \BibitemOpen
  \bibfield  {author} {\bibinfo {author} {\bibfnamefont {J.~J.}\ \bibnamefont
  {Madrid}}\ and\ \bibinfo {author} {\bibfnamefont {A.~M.}\ \bibnamefont
  {Mancho}},\ }\href@noop {} {\bibfield  {journal} {\bibinfo  {journal}
  {Chaos}\ }\textbf {\bibinfo {volume} {\textbf{19}}},\ \bibinfo {pages}
  {013111} (\bibinfo {year} {2009})}\BibitemShut {NoStop}%
\bibitem [{\citenamefont {Mancho}\ \emph {et~al.}(2013)\citenamefont {Mancho},
  \citenamefont {Wiggins}, \citenamefont {Curbelo},\ and\ \citenamefont
  {Mendoza}}]{MWCM13}%
  \BibitemOpen
  \bibfield  {author} {\bibinfo {author} {\bibfnamefont {A.~M.}\ \bibnamefont
  {Mancho}}, \bibinfo {author} {\bibfnamefont {S.}~\bibnamefont {Wiggins}},
  \bibinfo {author} {\bibfnamefont {J.}~\bibnamefont {Curbelo}},\ and\ \bibinfo
  {author} {\bibfnamefont {C.}~\bibnamefont {Mendoza}},\ }\href@noop {}
  {\bibfield  {journal} {\bibinfo  {journal} {Communications in Nonlinear
  Science and Numerical Simulation}\ }\textbf {\bibinfo {volume}
  {\textbf{18}}},\ \bibinfo {pages} {3530} (\bibinfo {year}
  {2013})}\BibitemShut {NoStop}%
\bibitem [{\citenamefont {Agaoglou}\ \emph {et~al.}(2020)\citenamefont
  {Agaoglou}, \citenamefont {Aguilar-Sanjuan}, \citenamefont
  {Garc{\'i}a-Garrido}, \citenamefont {Gonz{\'a}lez-Montoya}, \citenamefont
  {Katsanikas}, \citenamefont {Krajňák}, \citenamefont {Naik},\ and\
  \citenamefont {Wiggins}}]{AAGGKKNW20}%
  \BibitemOpen
  \bibfield  {author} {\bibinfo {author} {\bibfnamefont {M.}~\bibnamefont
  {Agaoglou}}, \bibinfo {author} {\bibfnamefont {B.}~\bibnamefont
  {Aguilar-Sanjuan}}, \bibinfo {author} {\bibfnamefont {V.~J.}\ \bibnamefont
  {Garc{\'i}a-Garrido}}, \bibinfo {author} {\bibfnamefont {F.}~\bibnamefont
  {Gonz{\'a}lez-Montoya}}, \bibinfo {author} {\bibfnamefont {M.}~\bibnamefont
  {Katsanikas}}, \bibinfo {author} {\bibfnamefont {V.}~\bibnamefont
  {Krajňák}}, \bibinfo {author} {\bibfnamefont {S.}~\bibnamefont {Naik}},\
  and\ \bibinfo {author} {\bibfnamefont {S.}~\bibnamefont {Wiggins}},\
  }\href@noop {} {\emph {\bibinfo {title} {Lagrangian Descriptors: {D}iscovery
  and Quantification of Phase Space Structure and Transport}}}\ (\bibinfo
  {publisher} {zenodo: 10.5281/zenodo.3958985},\ \bibinfo {year}
  {2020})\BibitemShut {NoStop}%
\bibitem [{\citenamefont {Craven}\ and\ \citenamefont
  {Hernandez}(2015)}]{CH15}%
  \BibitemOpen
  \bibfield  {author} {\bibinfo {author} {\bibfnamefont {G.~T.}\ \bibnamefont
  {Craven}}\ and\ \bibinfo {author} {\bibfnamefont {R.}~\bibnamefont
  {Hernandez}},\ }\href@noop {} {\bibfield  {journal} {\bibinfo  {journal}
  {Physical Review Letters}\ }\textbf {\bibinfo {volume} {\textbf{115}}},\
  \bibinfo {pages} {148301} (\bibinfo {year} {2015})}\BibitemShut {NoStop}%
\bibitem [{\citenamefont {Junginger}\ and\ \citenamefont
  {Hernandez}(2016)}]{JH16}%
  \BibitemOpen
  \bibfield  {author} {\bibinfo {author} {\bibfnamefont {A.}~\bibnamefont
  {Junginger}}\ and\ \bibinfo {author} {\bibfnamefont {R.}~\bibnamefont
  {Hernandez}},\ }\href@noop {} {\bibfield  {journal} {\bibinfo  {journal} {The
  Journal of Physical Chemistry B}\ }\textbf {\bibinfo {volume}
  {\textbf{120}}},\ \bibinfo {pages} {1720} (\bibinfo {year}
  {2016})}\BibitemShut {NoStop}%
\bibitem [{\citenamefont {Craven}\ \emph {et~al.}(2017)\citenamefont {Craven},
  \citenamefont {Junginger},\ and\ \citenamefont {Hernandez}}]{CJH17}%
  \BibitemOpen
  \bibfield  {author} {\bibinfo {author} {\bibfnamefont {G.~T.}\ \bibnamefont
  {Craven}}, \bibinfo {author} {\bibfnamefont {A.}~\bibnamefont {Junginger}},\
  and\ \bibinfo {author} {\bibfnamefont {R.}~\bibnamefont {Hernandez}},\
  }\href@noop {} {\bibfield  {journal} {\bibinfo  {journal} {Physical Review
  E}\ }\textbf {\bibinfo {volume} {\textbf{96}}},\ \bibinfo {pages} {022222}
  (\bibinfo {year} {2017})}\BibitemShut {NoStop}%
\bibitem [{\citenamefont {Revuelta}\ \emph {et~al.}(2019)\citenamefont
  {Revuelta}, \citenamefont {Benito},\ and\ \citenamefont {Borondo}}]{RBB19}%
  \BibitemOpen
  \bibfield  {author} {\bibinfo {author} {\bibfnamefont {F.}~\bibnamefont
  {Revuelta}}, \bibinfo {author} {\bibfnamefont {R.}~\bibnamefont {Benito}},\
  and\ \bibinfo {author} {\bibfnamefont {F.}~\bibnamefont {Borondo}},\
  }\href@noop {} {\bibfield  {journal} {\bibinfo  {journal} {Physical Review
  E}\ }\textbf {\bibinfo {volume} {\textbf{99}}},\ \bibinfo {pages} {032221}
  (\bibinfo {year} {2019})}\BibitemShut {NoStop}%
\bibitem [{\citenamefont {Katsanikas}\ \emph
  {et~al.}(2020{\natexlab{a}})\citenamefont {Katsanikas}, \citenamefont
  {Garc{\'\i}a-Garrido},\ and\ \citenamefont {Wiggins}}]{KGW20}%
  \BibitemOpen
  \bibfield  {author} {\bibinfo {author} {\bibfnamefont {M.}~\bibnamefont
  {Katsanikas}}, \bibinfo {author} {\bibfnamefont {V.}~\bibnamefont
  {Garc{\'\i}a-Garrido}},\ and\ \bibinfo {author} {\bibfnamefont
  {S.}~\bibnamefont {Wiggins}},\ }\href@noop {} {\bibfield  {journal} {\bibinfo
   {journal} {International Journal of Bifurcation and Chaos}\ }\textbf
  {\bibinfo {volume} {\textbf{30}}},\ \bibinfo {pages} {2030026} (\bibinfo
  {year} {2020}{\natexlab{a}})}\BibitemShut {NoStop}%
\bibitem [{\citenamefont {Agaoglou}\ \emph {et~al.}(2021)\citenamefont
  {Agaoglou}, \citenamefont {Garc{\'i}a-Garrido}, \citenamefont {Katsanikas},\
  and\ \citenamefont {Wiggins}}]{AGKW21}%
  \BibitemOpen
  \bibfield  {author} {\bibinfo {author} {\bibfnamefont {M.}~\bibnamefont
  {Agaoglou}}, \bibinfo {author} {\bibfnamefont {V.}~\bibnamefont
  {Garc{\'i}a-Garrido}}, \bibinfo {author} {\bibfnamefont {M.}~\bibnamefont
  {Katsanikas}},\ and\ \bibinfo {author} {\bibfnamefont {S.}~\bibnamefont
  {Wiggins}},\ }\href@noop {} {\bibfield  {journal} {\bibinfo  {journal}
  {Communications in Nonlinear Science and Numerical Simulation}\ }\textbf
  {\bibinfo {volume} {\textbf{103}}},\ \bibinfo {pages} {105993} (\bibinfo
  {year} {2021})}\BibitemShut {NoStop}%
\bibitem [{\citenamefont {Revuelta}\ \emph {et~al.}(2023)\citenamefont
  {Revuelta}, \citenamefont {Arranz}, \citenamefont {Benito},\ and\
  \citenamefont {Borondo}}]{RABB23}%
  \BibitemOpen
  \bibfield  {author} {\bibinfo {author} {\bibfnamefont {F.}~\bibnamefont
  {Revuelta}}, \bibinfo {author} {\bibfnamefont {F.}~\bibnamefont {Arranz}},
  \bibinfo {author} {\bibfnamefont {R.}~\bibnamefont {Benito}},\ and\ \bibinfo
  {author} {\bibfnamefont {F.}~\bibnamefont {Borondo}},\ }\href@noop {}
  {\bibfield  {journal} {\bibinfo  {journal} {Communications in Nonlinear
  Science and Numerical Simulation}\ }\textbf {\bibinfo {volume}
  {\textbf{123}}},\ \bibinfo {pages} {107265} (\bibinfo {year}
  {2023})}\BibitemShut {NoStop}%
\bibitem [{\citenamefont {Katsanikas}\ \emph {et~al.}(2023)\citenamefont
  {Katsanikas}, \citenamefont {Hillebrand}, \citenamefont {Skokos},\ and\
  \citenamefont {Wiggins}}]{KHSW23}%
  \BibitemOpen
  \bibfield  {author} {\bibinfo {author} {\bibfnamefont {M.}~\bibnamefont
  {Katsanikas}}, \bibinfo {author} {\bibfnamefont {M.}~\bibnamefont
  {Hillebrand}}, \bibinfo {author} {\bibfnamefont {C.}~\bibnamefont {Skokos}},\
  and\ \bibinfo {author} {\bibfnamefont {S.}~\bibnamefont {Wiggins}},\
  }\href@noop {} {\bibfield  {journal} {\bibinfo  {journal} {Chemical Physics
  Letters}\ }\textbf {\bibinfo {volume} {\textbf{811}}},\ \bibinfo {pages}
  {140208} (\bibinfo {year} {2023})}\BibitemShut {NoStop}%
\bibitem [{\citenamefont {Mendoza}\ and\ \citenamefont {Mancho}(2010)}]{MM10}%
  \BibitemOpen
  \bibfield  {author} {\bibinfo {author} {\bibfnamefont {C.}~\bibnamefont
  {Mendoza}}\ and\ \bibinfo {author} {\bibfnamefont {A.}~\bibnamefont
  {Mancho}},\ }\href@noop {} {\bibfield  {journal} {\bibinfo  {journal}
  {Physical Review Letters}\ }\textbf {\bibinfo {volume} {\textbf{105}}},\
  \bibinfo {pages} {038501} (\bibinfo {year} {2010})}\BibitemShut {NoStop}%
\bibitem [{\citenamefont {Garc{\'i}a-Garrido}\ \emph
  {et~al.}(2016)\citenamefont {Garc{\'i}a-Garrido}, \citenamefont {Ramos},
  \citenamefont {Mancho}, \citenamefont {Coca},\ and\ \citenamefont
  {Wiggins}}]{GRMCW16}%
  \BibitemOpen
  \bibfield  {author} {\bibinfo {author} {\bibfnamefont {V.}~\bibnamefont
  {Garc{\'i}a-Garrido}}, \bibinfo {author} {\bibfnamefont {A.}~\bibnamefont
  {Ramos}}, \bibinfo {author} {\bibfnamefont {A.~M.}\ \bibnamefont {Mancho}},
  \bibinfo {author} {\bibfnamefont {J.}~\bibnamefont {Coca}},\ and\ \bibinfo
  {author} {\bibfnamefont {S.}~\bibnamefont {Wiggins}},\ }\href@noop {}
  {\bibfield  {journal} {\bibinfo  {journal} {Marine Pollution Bulletin}\
  }\textbf {\bibinfo {volume} {\textbf{112}}},\ \bibinfo {pages} {201}
  (\bibinfo {year} {2016})}\BibitemShut {NoStop}%
\bibitem [{\citenamefont {Bruera}\ \emph {et~al.}(2023)\citenamefont {Bruera},
  \citenamefont {Curbelo}, \citenamefont {Garc{\'i}a-S{\'a}nchez},\ and\
  \citenamefont {Mancho}}]{BCGM23}%
  \BibitemOpen
  \bibfield  {author} {\bibinfo {author} {\bibfnamefont {R.}~\bibnamefont
  {Bruera}}, \bibinfo {author} {\bibfnamefont {J.}~\bibnamefont {Curbelo}},
  \bibinfo {author} {\bibfnamefont {G.}~\bibnamefont
  {Garc{\'i}a-S{\'a}nchez}},\ and\ \bibinfo {author} {\bibfnamefont
  {A.}~\bibnamefont {Mancho}},\ }\href@noop {} {\bibfield  {journal} {\bibinfo
  {journal} {Geophysical Research Letters}\ }\textbf {\bibinfo {volume}
  {\textbf{50}}},\ \bibinfo {pages} {e2022GL102244} (\bibinfo {year}
  {2023})}\BibitemShut {NoStop}%
\bibitem [{\citenamefont {Darwish}\ \emph {et~al.}(2021)\citenamefont
  {Darwish}, \citenamefont {Norouzi}, \citenamefont {Di~Labbio},\ and\
  \citenamefont {Kadem}}]{DNDK21}%
  \BibitemOpen
  \bibfield  {author} {\bibinfo {author} {\bibfnamefont {A.}~\bibnamefont
  {Darwish}}, \bibinfo {author} {\bibfnamefont {S.}~\bibnamefont {Norouzi}},
  \bibinfo {author} {\bibfnamefont {G.}~\bibnamefont {Di~Labbio}},\ and\
  \bibinfo {author} {\bibfnamefont {L.}~\bibnamefont {Kadem}},\ }\href@noop {}
  {\bibfield  {journal} {\bibinfo  {journal} {Physics of Fluids}\ }\textbf
  {\bibinfo {volume} {\textbf{33}}},\ \bibinfo {pages} {111707} (\bibinfo
  {year} {2021})}\BibitemShut {NoStop}%
\bibitem [{\citenamefont {Amahjour}\ \emph {et~al.}(2023)\citenamefont
  {Amahjour}, \citenamefont {Garc{\'i}a-Sánchez}, \citenamefont {Agaoglou},\
  and\ \citenamefont {Mancho}}]{AGAM23}%
  \BibitemOpen
  \bibfield  {author} {\bibinfo {author} {\bibfnamefont {N.}~\bibnamefont
  {Amahjour}}, \bibinfo {author} {\bibfnamefont {G.}~\bibnamefont
  {Garc{\'i}a-Sánchez}}, \bibinfo {author} {\bibfnamefont {M.}~\bibnamefont
  {Agaoglou}},\ and\ \bibinfo {author} {\bibfnamefont {A.~M.}\ \bibnamefont
  {Mancho}},\ }\href@noop {} {\bibfield  {journal} {\bibinfo  {journal}
  {Physica D}\ }\textbf {\bibinfo {volume} {\textbf{453}}},\ \bibinfo {pages}
  {133825} (\bibinfo {year} {2023})}\BibitemShut {NoStop}%
\bibitem [{\citenamefont {Daquin}\ \emph {et~al.}(2022)\citenamefont {Daquin},
  \citenamefont {P\'edenon-Orlanducci}, \citenamefont {Agaoglou}, \citenamefont
  {Garc\'ia-S\'anchez},\ and\ \citenamefont {Mancho}}]{DPAGM22}%
  \BibitemOpen
  \bibfield  {author} {\bibinfo {author} {\bibfnamefont {J.}~\bibnamefont
  {Daquin}}, \bibinfo {author} {\bibfnamefont {R.}~\bibnamefont
  {P\'edenon-Orlanducci}}, \bibinfo {author} {\bibfnamefont {M.}~\bibnamefont
  {Agaoglou}}, \bibinfo {author} {\bibfnamefont {G.}~\bibnamefont
  {Garc\'ia-S\'anchez}},\ and\ \bibinfo {author} {\bibfnamefont {A.~M.}\
  \bibnamefont {Mancho}},\ }\href@noop {} {\bibfield  {journal} {\bibinfo
  {journal} {Physica D}\ }\textbf {\bibinfo {volume} {\textbf{442}}},\ \bibinfo
  {pages} {133520} (\bibinfo {year} {2022})}\BibitemShut {NoStop}%
\bibitem [{\citenamefont {Hillebrand}\ \emph {et~al.}(2022)\citenamefont
  {Hillebrand}, \citenamefont {Zimper}, \citenamefont {Ngapasare},
  \citenamefont {Katsanikas}, \citenamefont {Wiggins},\ and\ \citenamefont
  {Skokos}}]{HZNKWS22}%
  \BibitemOpen
  \bibfield  {author} {\bibinfo {author} {\bibfnamefont {M.}~\bibnamefont
  {Hillebrand}}, \bibinfo {author} {\bibfnamefont {S.}~\bibnamefont {Zimper}},
  \bibinfo {author} {\bibfnamefont {A.}~\bibnamefont {Ngapasare}}, \bibinfo
  {author} {\bibfnamefont {M.}~\bibnamefont {Katsanikas}}, \bibinfo {author}
  {\bibfnamefont {S.}~\bibnamefont {Wiggins}},\ and\ \bibinfo {author}
  {\bibfnamefont {C.}~\bibnamefont {Skokos}},\ }\href@noop {} {\bibfield
  {journal} {\bibinfo  {journal} {Chaos}\ }\textbf {\bibinfo {volume} {32}},\
  \bibinfo {pages} {\textbf{123122}} (\bibinfo {year} {2022})}\BibitemShut
  {NoStop}%
\bibitem [{\citenamefont {Zimper}\ \emph {et~al.}(2023)\citenamefont {Zimper},
  \citenamefont {Ngapasare}, \citenamefont {Hillebrand}, \citenamefont
  {Katsanikas}, \citenamefont {Wiggins},\ and\ \citenamefont
  {Skokos}}]{ZNHKWS23}%
  \BibitemOpen
  \bibfield  {author} {\bibinfo {author} {\bibfnamefont {S.}~\bibnamefont
  {Zimper}}, \bibinfo {author} {\bibfnamefont {A.}~\bibnamefont {Ngapasare}},
  \bibinfo {author} {\bibfnamefont {M.}~\bibnamefont {Hillebrand}}, \bibinfo
  {author} {\bibfnamefont {M.}~\bibnamefont {Katsanikas}}, \bibinfo {author}
  {\bibfnamefont {S.}~\bibnamefont {Wiggins}},\ and\ \bibinfo {author}
  {\bibfnamefont {C.}~\bibnamefont {Skokos}},\ }\href@noop {} {\bibfield
  {journal} {\bibinfo  {journal} {Physica D}\ }\textbf {\bibinfo {volume}
  {\textbf{453}}},\ \bibinfo {pages} {133833} (\bibinfo {year}
  {2023})}\BibitemShut {NoStop}%
\bibitem [{\citenamefont {Daquin}\ and\ \citenamefont
  {Charalambous}(2023)}]{DC23}%
  \BibitemOpen
  \bibfield  {author} {\bibinfo {author} {\bibfnamefont {J.}~\bibnamefont
  {Daquin}}\ and\ \bibinfo {author} {\bibfnamefont {C.}~\bibnamefont
  {Charalambous}},\ }\href@noop {} {\bibfield  {journal} {\bibinfo  {journal}
  {Celestial Mechanics and Dynamical Astronomy}\ }\textbf {\bibinfo {volume}
  {\textbf{135}}},\ \bibinfo {pages} {31} (\bibinfo {year} {2023})}\BibitemShut
  {NoStop}%
\bibitem [{\citenamefont {Căliman}\ \emph {et~al.}(2025)\citenamefont
  {Căliman}, \citenamefont {Daquin},\ and\ \citenamefont {Libert}}]{CDL25}%
  \BibitemOpen
  \bibfield  {author} {\bibinfo {author} {\bibfnamefont {A.}~\bibnamefont
  {Căliman}}, \bibinfo {author} {\bibfnamefont {J.}~\bibnamefont {Daquin}},\
  and\ \bibinfo {author} {\bibfnamefont {A.-S.}\ \bibnamefont {Libert}},\
  }\href@noop {} {\bibfield  {journal} {\bibinfo  {journal} {Physica D}\
  }\textbf {\bibinfo {volume} {\textbf{472}}},\ \bibinfo {pages} {134506}
  (\bibinfo {year} {2025})}\BibitemShut {NoStop}%
\bibitem [{\citenamefont {Katsanikas}\ and\ \citenamefont
  {Patsis}(2011)}]{KP11}%
  \BibitemOpen
  \bibfield  {author} {\bibinfo {author} {\bibfnamefont {M.}~\bibnamefont
  {Katsanikas}}\ and\ \bibinfo {author} {\bibfnamefont {P.}~\bibnamefont
  {Patsis}},\ }\href@noop {} {\bibfield  {journal} {\bibinfo  {journal}
  {International Journal of Bifurcation and Chaos}\ }\textbf {\bibinfo {volume}
  {\textbf{21}}},\ \bibinfo {pages} {467} (\bibinfo {year} {2011})}\BibitemShut
  {NoStop}%
\bibitem [{\citenamefont {Katsanikas}\ \emph {et~al.}(2011)\citenamefont
  {Katsanikas}, \citenamefont {Patsis},\ and\ \citenamefont
  {Contopoulos}}]{KPC11}%
  \BibitemOpen
  \bibfield  {author} {\bibinfo {author} {\bibfnamefont {M.}~\bibnamefont
  {Katsanikas}}, \bibinfo {author} {\bibfnamefont {P.}~\bibnamefont {Patsis}},\
  and\ \bibinfo {author} {\bibfnamefont {G.}~\bibnamefont {Contopoulos}},\
  }\href@noop {} {\bibfield  {journal} {\bibinfo  {journal} {International
  Journal of Bifurcation and Chaos}\ }\textbf {\bibinfo {volume}
  {\textbf{21}}},\ \bibinfo {pages} {2321} (\bibinfo {year}
  {2011})}\BibitemShut {NoStop}%
\bibitem [{\citenamefont {Katsanikas}\ \emph {et~al.}(2013)\citenamefont
  {Katsanikas}, \citenamefont {Patsis},\ and\ \citenamefont
  {Contopoulos}}]{KPC13}%
  \BibitemOpen
  \bibfield  {author} {\bibinfo {author} {\bibfnamefont {M.}~\bibnamefont
  {Katsanikas}}, \bibinfo {author} {\bibfnamefont {P.}~\bibnamefont {Patsis}},\
  and\ \bibinfo {author} {\bibfnamefont {G.}~\bibnamefont {Contopoulos}},\
  }\href@noop {} {\bibfield  {journal} {\bibinfo  {journal} {International
  Journal of Bifurcation and Chaos}\ }\textbf {\bibinfo {volume}
  {\textbf{23}}},\ \bibinfo {pages} {1330005} (\bibinfo {year}
  {2013})}\BibitemShut {NoStop}%
\bibitem [{\citenamefont {Hillebrand}\ \emph {et~al.}(2023)\citenamefont
  {Hillebrand}, \citenamefont {Katsanikas}, \citenamefont {Wiggins},\ and\
  \citenamefont {Skokos}}]{HKWS23}%
  \BibitemOpen
  \bibfield  {author} {\bibinfo {author} {\bibfnamefont {M.}~\bibnamefont
  {Hillebrand}}, \bibinfo {author} {\bibfnamefont {M.}~\bibnamefont
  {Katsanikas}}, \bibinfo {author} {\bibfnamefont {S.}~\bibnamefont
  {Wiggins}},\ and\ \bibinfo {author} {\bibfnamefont {C.}~\bibnamefont
  {Skokos}},\ }\href@noop {} {\bibfield  {journal} {\bibinfo  {journal}
  {Physical Review E}\ }\textbf {\bibinfo {volume} {\textbf{108}}},\ \bibinfo
  {pages} {024211} (\bibinfo {year} {2023})}\BibitemShut {NoStop}%
\bibitem [{\citenamefont {Miyamoto}\ and\ \citenamefont {Nagai}(1975)}]{MN75}%
  \BibitemOpen
  \bibfield  {author} {\bibinfo {author} {\bibfnamefont {M.}~\bibnamefont
  {Miyamoto}}\ and\ \bibinfo {author} {\bibfnamefont {R.}~\bibnamefont
  {Nagai}},\ }\href@noop {} {\bibfield  {journal} {\bibinfo  {journal}
  {Publications of the Astronomical Society of Japan}\ }\textbf {\bibinfo
  {volume} {\textbf{27}}},\ \bibinfo {pages} {533} (\bibinfo {year}
  {1975})}\BibitemShut {NoStop}%
\bibitem [{\citenamefont {Danby}(1965)}]{D65}%
  \BibitemOpen
  \bibfield  {author} {\bibinfo {author} {\bibfnamefont {J.}~\bibnamefont
  {Danby}},\ }\href@noop {} {\bibfield  {journal} {\bibinfo  {journal} {The
  Astronomical Journal}\ }\textbf {\bibinfo {volume} {\textbf{70}}},\ \bibinfo
  {pages} {501} (\bibinfo {year} {1965})}\BibitemShut {NoStop}%
\bibitem [{\citenamefont {Contopoulos}(1990)}]{C90}%
  \BibitemOpen
  \bibfield  {author} {\bibinfo {author} {\bibfnamefont {G.}~\bibnamefont
  {Contopoulos}},\ }\href@noop {} {\bibfield  {journal} {\bibinfo  {journal}
  {Astronomy and Astrophysics}\ }\textbf {\bibinfo {volume} {\textbf{231}}},\
  \bibinfo {pages} {41} (\bibinfo {year} {1990})}\BibitemShut {NoStop}%
\bibitem [{\citenamefont {Koon}\ \emph
  {et~al.}(2001{\natexlab{a}})\citenamefont {Koon}, \citenamefont {Lo},
  \citenamefont {Marsden},\ and\ \citenamefont {Ross}}]{KLMR01a}%
  \BibitemOpen
  \bibfield  {author} {\bibinfo {author} {\bibfnamefont {W.}~\bibnamefont
  {Koon}}, \bibinfo {author} {\bibfnamefont {M.}~\bibnamefont {Lo}}, \bibinfo
  {author} {\bibfnamefont {J.}~\bibnamefont {Marsden}},\ and\ \bibinfo {author}
  {\bibfnamefont {S.}~\bibnamefont {Ross}},\ }\href@noop {} {\bibfield
  {journal} {\bibinfo  {journal} {Celestial Mechanics and Dynamical Astronomy}\
  }\textbf {\bibinfo {volume} {\textbf{81}}},\ \bibinfo {pages} {27} (\bibinfo
  {year} {2001}{\natexlab{a}})}\BibitemShut {NoStop}%
\bibitem [{\citenamefont {Koon}\ \emph
  {et~al.}(2001{\natexlab{b}})\citenamefont {Koon}, \citenamefont {Lo},
  \citenamefont {Marsden},\ and\ \citenamefont {Ross}}]{KLMR01b}%
  \BibitemOpen
  \bibfield  {author} {\bibinfo {author} {\bibfnamefont {W.}~\bibnamefont
  {Koon}}, \bibinfo {author} {\bibfnamefont {M.}~\bibnamefont {Lo}}, \bibinfo
  {author} {\bibfnamefont {J.}~\bibnamefont {Marsden}},\ and\ \bibinfo {author}
  {\bibfnamefont {S.}~\bibnamefont {Ross}},\ }\href@noop {} {\bibfield
  {journal} {\bibinfo  {journal} {Celestial Mechanics and Dynamical Astronomy}\
  }\textbf {\bibinfo {volume} {\textbf{81}}},\ \bibinfo {pages} {63} (\bibinfo
  {year} {2001}{\natexlab{b}})}\BibitemShut {NoStop}%
\bibitem [{\citenamefont {Luther}(1968)}]{L68}%
  \BibitemOpen
  \bibfield  {author} {\bibinfo {author} {\bibfnamefont {H.}~\bibnamefont
  {Luther}},\ }\href@noop {} {\bibfield  {journal} {\bibinfo  {journal}
  {Mathematics of Computation}\ }\textbf {\bibinfo {volume} {\textbf{22}}},\
  \bibinfo {pages} {434} (\bibinfo {year} {1968})}\BibitemShut {NoStop}%
\bibitem [{\citenamefont {Lopesino}\ \emph {et~al.}(2017)\citenamefont
  {Lopesino}, \citenamefont {Balibrea}, \citenamefont {Garcia-Garrido},
  \citenamefont {Wiggins},\ and\ \citenamefont {Mancho}}]{LBGWM17}%
  \BibitemOpen
  \bibfield  {author} {\bibinfo {author} {\bibfnamefont {C.}~\bibnamefont
  {Lopesino}}, \bibinfo {author} {\bibfnamefont {F.}~\bibnamefont {Balibrea}},
  \bibinfo {author} {\bibfnamefont {V.}~\bibnamefont {Garcia-Garrido}},
  \bibinfo {author} {\bibfnamefont {S.}~\bibnamefont {Wiggins}},\ and\ \bibinfo
  {author} {\bibfnamefont {A.}~\bibnamefont {Mancho}},\ }\href@noop {}
  {\bibfield  {journal} {\bibinfo  {journal} {International Journal of
  Bifurcation and Chaos}\ }\textbf {\bibinfo {volume} {\textbf{27}}},\ \bibinfo
  {pages} {1730001} (\bibinfo {year} {2017})}\BibitemShut {NoStop}%
\bibitem [{\citenamefont {Demian}\ and\ \citenamefont {Wiggins}(2017)}]{DW17}%
  \BibitemOpen
  \bibfield  {author} {\bibinfo {author} {\bibfnamefont {A.}~\bibnamefont
  {Demian}}\ and\ \bibinfo {author} {\bibfnamefont {S.}~\bibnamefont
  {Wiggins}},\ }\href@noop {} {\bibfield  {journal} {\bibinfo  {journal}
  {International Journal of Bifurcation and Chaos}\ }\textbf {\bibinfo {volume}
  {\textbf{27}}},\ \bibinfo {pages} {1750225} (\bibinfo {year}
  {2017})}\BibitemShut {NoStop}%
\bibitem [{\citenamefont {Katsanikas}\ \emph
  {et~al.}(2020{\natexlab{b}})\citenamefont {Katsanikas}, \citenamefont
  {Garc{\'{\i}}a-Garrido}, \citenamefont {Agaoglou},\ and\ \citenamefont
  {Wiggins}}]{KGAW20}%
  \BibitemOpen
  \bibfield  {author} {\bibinfo {author} {\bibfnamefont {M.}~\bibnamefont
  {Katsanikas}}, \bibinfo {author} {\bibfnamefont {V.~J.}\ \bibnamefont
  {Garc{\'{\i}}a-Garrido}}, \bibinfo {author} {\bibfnamefont {M.}~\bibnamefont
  {Agaoglou}},\ and\ \bibinfo {author} {\bibfnamefont {S.}~\bibnamefont
  {Wiggins}},\ }\href@noop {} {\bibfield  {journal} {\bibinfo  {journal}
  {Physical Review E}\ }\textbf {\bibinfo {volume} {\textbf{102}}},\ \bibinfo
  {pages} {012215} (\bibinfo {year} {2020}{\natexlab{b}})}\BibitemShut
  {NoStop}%
\end{thebibliography}%

\end{document}